\documentclass[aps,prb,reprint,twocolumn,notitlepage,floatfix,tightenlines,superscriptaddress,longbibliography]{revtex4-2}
\usepackage[latin9]{inputenc}
\usepackage{natbib}
\usepackage{amsfonts}
\usepackage{amsmath}
\usepackage{amssymb}
\usepackage[T1]{fontenc}
\usepackage{graphicx}

\makeatletter

\numberwithin{equation}{section}

\usepackage[colorlinks=true,citecolor=blue,linkcolor=blue,urlcolor=blue]{hyperref}

\makeatother

\newcommand{\vect}[1] {\mathbf{#1}}   % defining the notation for a vector in this papar
\newcommand{\tens}[1] {\tensor{#1}}   % tensor, defined in revsymb.sty

\renewcommand{\vec}[1]{\mathbf{#1}}
\newcommand{\intdkdw}{\int \frac{d\vec{k}}{(2\pi)^2} \int \frac{d\omega}{2\pi}}
\newcommand{\intdkodw}{\int \frac{d\vec{k}}{(2\pi)^2} \oint \frac{d\omega}{2\pi}}
\newcommand{\intdk}{\int \frac{d\vec{k}}{(2\pi)^2}}
\newcommand{\intdw}{\int \frac{d\omega}{2\pi}}
\newcommand{\ointdw}{\oint \frac{d\omega}{2\pi}}
\newcommand{\Tr}{\mathrm{Tr}}
\newcommand{\R}{\mathrm{R}}
\newcommand{\A}{\mathrm{A}}

\newcommand{\ImSigma}{\mathrm{Im}\Sigma}
\newcommand{\ReSigma}{\mathrm{Re}\Sigma}
\newcommand{\Glesser}{G^{<}}
\newcommand{\vecx}{\vec{x}}
\newcommand{\vecxp}{\vec{x}^\prime}
\newcommand{\xt} {\vect{x}t}
\newcommand{\xptp}{\vec{x}^\prime t^\prime}
\newcommand{\dkdw}{\frac{d\vec{k}}{(2\pi)^2} \frac{d\omega}{2\pi}}
\newcommand{\gradT}{\vect{\Theta}}
\newcommand{\Li}{\mathrm{Li}}

\newcommand{\HI}{H_{\mathrm{I}}}
\newcommand{\HII}{H_{\mathrm{II}}}

\newcommand{\soc}{\lambda_{\text{soc}} }
\newcommand{\dif} {\mathrm{d}}  % differential operator, distinguished from the dimension $d$
\newcommand{\evolve} {\frac{\partial}{\partial t}}

\def\tr{\mathrm{tr}}

\begin{document}

\title{Heat-bath approach to anomalous thermal transport: effects of inelastic scattering}

\author{Zhiqiang Wang}
\email[]{zqwang@uchicago.edu}
\affiliation{Department of Physics and James Franck Institute, University of Chicago, Chicago, Illinois 60637, USA}
\author{Rufus Boyack}
\affiliation{D\'epartement de physique, Universit\'e de Montr\'eal, Montr\'eal, Qu\'ebec H3C 3J7, Canada}
\author{K. Levin}
\email[]{levin@jfi.uchicago.edu}
\affiliation{Department of Physics and James Franck Institute, University of Chicago, Chicago, Illinois 60637, USA}

\date{\today}
\begin{abstract}
We present results for the entire set of anomalous charge and heat transport coefficients for metallic systems in the presence of a finite-temperature heat bath. In realistic physical systems
this necessitates the inclusion of
inelastic dissipation mechanisms; relatively little is known theoretically
about their effects on anomalous transport.
Here we demonstrate 
how these dissipative processes are strongly intertwined with Berry-curvature physics.
Our calculations are made possible by the introduction of a Caldeira-Leggett reservoir which allows us to
avoid the sometimes-problematic device
of the pseudogravitational potential.
Using our formulas, we focus on the finite-temperature behavior of the important
anomalous Wiedemann-Franz ratio.
Despite previous expectations, this ratio is found to be non-universal
as it can exhibit either an upturn or a downturn as
temperature increases away from zero. We emphasize that this derives from a \textit{competition} between Berry curvatures having different signs in different regions of the Brillouin zone.  
We point to experimental support for these observations and
for the behavior of an alternative ratio involving a thermoelectric response which, by contrast, appears to
be more universal at low temperatures. Our work paves the way for future theory and experiment, demonstrating 
how inelastic scattering at non-zero temperature affects the behavior of
all anomalous transport coefficients.
\end{abstract}

\maketitle

\section{Introduction}

Charge and thermal transport (both with and without an applied magnetic field)
are viewed as providing fundamental information about a condensed matter system.  
Our understanding of these response properties
has recently progressed largely because of the focus on
topological quantum materials. 
Initially, the electrical Hall conductivity was at the heart of this excitement,
but attention has now shifted to
its thermal analogue~\cite{Banerjee2018,Qin2011}--the thermal Hall conductivity-- 
and to anomalous counterparts of
the Nernst and Ettinghausen thermoelectric coefficients~\cite{Xiao2006}.
In addition to revelations about topological order~\cite{Kane1997,Cappelli2002,Levin2007,Lee2007,Banerjee2018}, 
in metallic systems this larger class of thermal and thermoelectric transport 
contains contributions which derive from Berry-curvature effects 
~\cite{Xiao2006,Xiao2010,Nagaosa2010} in the underlying band structure.
These appear even when the magnetic field is zero;
here the associated zero-field response coefficients constitute a class of so-called
``anomalous'' transport properties.

These anomalous transport coefficients are the focus of the present paper. 
They have been well-studied experimentally
over the entire range of temperatures~\cite{Shiomi2010, Li2017,Xu2020}.
Clearly, understanding these finite-temperature transport data requires addressing
inelastic dissipation effects, as phonons are always present. 
It is notable, then, that in the theoretical
literature the anomalous Berry-curvature contributions
have been restricted~\cite{Nagaosa2010} to low temperatures, where inelastic
processes can be ignored and special connections between the transverse anomalous
transport coefficients (such as the Wiedemann-Franz and Mott relations)
obtain~\cite{Qin2011,Xiao2006,Onoda2008}. 

This paper addresses this shortcoming. We establish what happens to the
anomalous charge and heat transport at finite temperatures when
inelastic dissipation plays a role, and we focus here on the entire set of transport coefficients. 
A key finding is our observation that
inelastic-scattering properties are strongly intertwined with Berry-curvature physics, 
as is concretely reflected in the behavior of the transverse Wiedemann-Franz (WF) ratio. 
We find that the presence of inelastic
dissipation can cause this ratio to exhibit an unexpected~\cite{Li2017,Xu2022}
upturn as temperature increases away from zero. 
This finding appears to be supported by experiment~\cite{Xu2022}. 
Central to our paper is the presentation of an alternative methodology for deriving
thermal transport coefficients which avoids many of the ambiguities and
pitfalls in more conventional transport schemes. 
We do so by the introduction of a heat bath such as considered by Caldeira and Leggett~\cite{Caldeira1983}.

Among the complications in understanding thermal transport is the fact
that an applied temperature gradient $\nabla T$ represents a ``statistical'' rather than a mechanical
force; this has led Luttinger~\cite{Luttinger1964} to develop a novel, albeit
unintuitive transport framework. Here
a ``gravitational pseudopotential'' is introduced which
gives rise to an effective mechanical force, in analogy with the electric
field $\mathbf{E}$ that drives charge transport. 
This approach builds on Tolman's earlier work~\cite{Tolman1930} demonstrating
that addressing transport in curved spacetime requires understanding
how to treat a position-dependent temperature $T(x)$. (Hereafter, we will omit the spatial dependence of the temperature $T(x)$.)
It has, however, been argued to
be problematic~\cite{Kapustin2020,Stone2012,Tatara2015,Gromov2015,Shitade2014,Park2021},
particularly for topological systems.
Indeed, lattice effects add an additional complexity, as in this case
for the gravitational analogue a temperature gradient 
would correspond to a distorted lattice with spatially dependent hopping.

A second challenge for computing transverse thermal and thermoelectric transport coefficients comes from the necessity~\cite{Cooper1997} 
to subtract magnetization heat and charge currents which do not contribute to bulk transport. 
This subtraction is crucial in order for the derived transport coefficients to be consistent with the third law
of thermodynamics and to satisfy the Onsager reciprocal relations~\cite{Cooper1997,Qin2011}. 
Although the spontaneous charge orbital magnetization~\footnote{We consider only orbital magnetizations in this paper and ignore any spin contributions.} 
in the absence of inelastic scattering is now relatively well established~\cite{Xiao2006,Shi2007,Bianco2013}, 
its heat counterpart is currently still being debated~\cite{Qin2011,Xiao2020,Kapustin2020}.
Moreover, much less is known about how to incorporate finite-lifetime effects when inelastic scattering is present.

\begin{figure}[tp]
\begin{center}
\includegraphics[width=0.75\linewidth, trim=0mm 5mm 0mm 0mm]{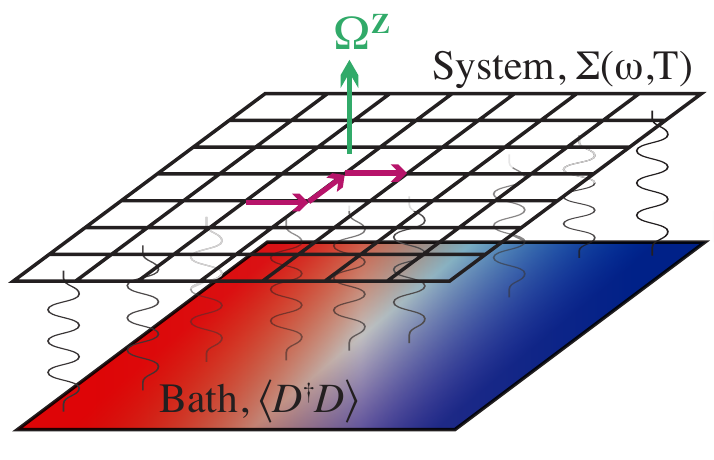}
\caption{Schematic diagram of our heat-bath approach. We consider a general multi-orbital lattice model of electrons
embedded in a heat bath, where the two components are in the same two-dimensional space even though in the diagram they are separated for better visualization. 
The figure represents the hopping of electrons under the influence of the $z$-component of Berry curvature (dark green vertical arrows with $\Omega^z$). 
The coupling of the electrons to the heat bath leads
to a nonzero dissipation characterized by the electronic self energy $\Sigma$.
The equilibrium of the system is maintained through the bath so that
a correlator for the bath particles, $\langle D^\dagger D\rangle$, is proportional to the
Fermi-Dirac distribution at a given local temperature. 
In thermal transport, a temperature gradient (represented by
color changes from red to blue) is applied across the sample to drive a heat current.}
\label{fig:Fig1}
\end{center}
\end{figure}

Here we consider a multi-orbital system of electrons confined to a lattice embedded in
a heat-bath reservoir which serves the dual purpose
of maintaining local equilibrium associated with the varying temperatures and establishing the temperature
gradients. This system is schematically represented in Fig.~\ref{fig:Fig1}. 
The reservoir consists of an infinite number of localized degrees of freedom (electrons in the present case) 
that can exchange energy and particles with the system. 
Importantly, a (bilinear) coupling between the system and bath particles gives rise to quantum dissipation, represented by a frequency and temperature-dependent electron self-energy, 
$\Sigma(\omega,T)$~\footnote{For brevity, we will sometimes suppress the $T$ dependence of $\Sigma$ and show it explicitly only when necessary.}. 
When inter-particle interactions between the system particles are absent (as we assume here) we
can \textit{exactly} derive all thermal and charge transport coefficients, in the presence of non-zero $\Sigma$.

A crucial, but subtle complication arises in thermal transport because of this inelastic dissipation.
In the presence of a temperature gradient and for
small temperature changes one might expect
contributions from the temperature derivative $\partial \Sigma /\partial T$
to enter into the calculated coefficients. 
That is, owing to the temperature gradient,
the particles, which are doing the thermal
conduction, experience a different dissipation at different points across the sample.
Because this is rather different from the behavior experienced in the presence of an electric field, it appears problematic. 
One can anticipate that, if such terms were present
in the response to a temperature gradient but not to an electric field,
they would compromise the Onsager reciprocal relations between electrical and thermal response. 

Indeed, we will show in this paper how
contributions from $\partial \Sigma/ \partial T$ are rather elegantly cancelled in both
the anomalous thermoelectric response coefficients,
as well as the anomalous thermal Hall conductivity.
We find this restoration of Onsager reciprocity relations is a natural consequence of the fluctuation-dissipation theorem~
\footnote{Note that this cancellation is rather non-trivial for the transverse transport coefficients, 
as the appropriate charge and heat magnetization currents have to be properly included in the calculations.}. 
We emphasize the fundamental role of this theorem throughout the paper noting that
within the ``gravitational-pseudopotential'' approach~\cite{Luttinger1964,Cooper1997}, 
the thermal Einstein relation presumably provides an analogous supporting framework.

Historically, concepts of fluctuations and stochastic processes, and the fluctuation-dissipation theorem, have played an
essential role in establishing not only Onsager's reciprocity but also in building the foundations of non-equilibrium thermodynamics~\cite{DeGroot1984}. 
Indeed, the fundamental Kadanoff-Baym transport equations~\cite{Kadanoff1962} have been shown to be
equivalent to a stochastic Langevin equation associated with a thermal bath~\cite{Anisimov2009, Greiner1998}.
This body of work has focused more on non-equilibrium quantum field theories in particle physics and cosmology rather than
in condensed matter systems, but the wider appeal of a more intuitive bath approach, such as
we apply here, should be evident.

The remainder of the paper is organized as follows. In Sec.~\ref{sec:general} we outline general procedures for
obtaining transport coefficients within our heat-bath approach and summarize the resulting central formulas. In Sec.~\ref{sec:Transport} we provide
a detailed derivation of the anomalous thermal Hall conductivity $\kappa_{xy}$ which incorporates the fundamental
fluctuation-dissipation relation. Numerical implications of our derived transport coefficient formulas are discussed in Sec.~\ref{sec:numerics}, 
emphasizing the interplay of Berry curvature and inelastic scattering. This is presented in
the context of the anomalous WF law as well as for an alternative thermoelectric transport ratio.
Section~\ref{sec:conclusion} contains our conclusions and outlook. 
In addition, we make a number of details available in several appendices.

\section{General discussion and summary of the formulas}
\label{sec:general}

\subsection{General Definitions}

Thermoelectric transport involves the response of charge (transport) current density $\vect{J}^e_{\tr}$
and heat (transport) current density $\vect{J}^h_{\tr}$ in the presence of an electric field $\vect{E}$ 
and temperature gradient $\vect{\Theta} \equiv - \nabla T$. 
We define four tensorial transport coefficients as : 
\begin{subequations} \label{eq:Currents}
\begin{align}
\begin{bmatrix}
\vect{J}_{\tr}^e \\
\vect{J}_{\tr}^h
\end{bmatrix} 
& \equiv 
\begin{bmatrix}
\tens{\sigma}     & \tens{\beta} \\
\tens{\gamma}   & \tens{\kappa}
\end{bmatrix}
\begin{bmatrix}
\vect{E}    \\
\vect{\Theta}
\end{bmatrix}  
= 
\begin{bmatrix}
\tens{\mathcal{L}}^{00}    & \tens{\mathcal{L}}^{01} \\
\tens{\mathcal{L}}^{10}   & \tens{\mathcal{L}}^{11}
\end{bmatrix}
\begin{bmatrix}
\vect{E}    \\
\vect{\Theta}
\end{bmatrix} 
\\
& = 
\begin{bmatrix}
\tens{L}^{e,e}  \quad  & \tens{L}^{e,h} - \frac{\partial \vect{M}^e }{\partial T} \times       \\
 \tens{L}^{h,e} - \vect{M}^e \times    \quad  &  \tens{L}^{h,h} - \frac{\partial \vect{M}^h }{\partial T} \times
\end{bmatrix}
\begin{bmatrix}
\vect{E}    \\
\vect{\Theta}
\end{bmatrix}.
\label{eq:Msubtraction}
\end{align}
\end{subequations}
Here, $\tens{\sigma}$ and $\tens{\kappa}$ are the electrical and thermal conductivities, respectively,
and $\tens{\gamma}$ and $\tens{\beta}$ represent the cross thermoelectric terms which are
related to one another by the Onsager reciprocal relation. 
In this paper we ignore the fact that in experiments $\tensor{\kappa}$ is different from the measured thermal conductivity due to an open-circuit
condition, which our calculations indicate will cause only a small correction to the anomalous Wiedemann-Franz ratio in the clean limit
\footnote{See Ref.~\cite{Boyack2021} and references therein for discussions in a different context.}.
For later convenience, we also introduce a more compact notation, $\tens{\mathcal{L}}^{\alpha \beta}$ with $\{\alpha,\beta\}=\{0,1\}$, for these coefficients. 

To calculate $\vect{J}^e_{\tr}$ and $\vect{J}^h_{\tr}$, one first computes the so-called ``microscopic'' current densities in the presence of external perturbations $\vec{E}$
and $\vect{\Theta}$, giving rise to the $\tens{L}^{ij}$ contribution in Eq.~\eqref{eq:Msubtraction}, with $\{i, j\}= \{ e, h\}$. 
It is well known that these ``microscopic'' currents are different from the transport currents ~\cite{Cooper1997,Qin2011}, 
and that to obtain the latter, physically measurable currents one needs to subtract the divergence-free currents 
due to the charge magnetization $\vect{M}^{e}$ and the heat magnetization $\vect{M}^{h}$. 
From Eq.~\eqref{eq:Msubtraction}, this subtraction is seen to affect only the transverse transport coefficients.

\subsection{Central Results}
\label{sec:summary}

In this paper we will show how both the coefficients $\tens{L}^{i,j}$ and the magnetization derivative $\partial \vec{M}^j/\partial T$ 
in Eq.~\eqref{eq:Msubtraction}  can be expressed in terms of the single-particle Green's function.
The Green's functions as specified include both
retarded
($G_{\R}$) and advanced
($G_{\A}$) 
forms which, importantly, depend on their respective self energies, $\Sigma_{\R}$ and $\Sigma_\A$.
We set $\hbar=k_{B}=a=1$ in the following, where $a$ is the lattice spacing.
% with $q=-e<0$ taken as the charge of electrons and
%$\beta=1/T$.
Throughout, the summation over repeated indices is assumed.

\begin{figure}[htp]
\begin{center}
\includegraphics[width=0.9\linewidth,trim=0mm 0mm 0mm 0mm]{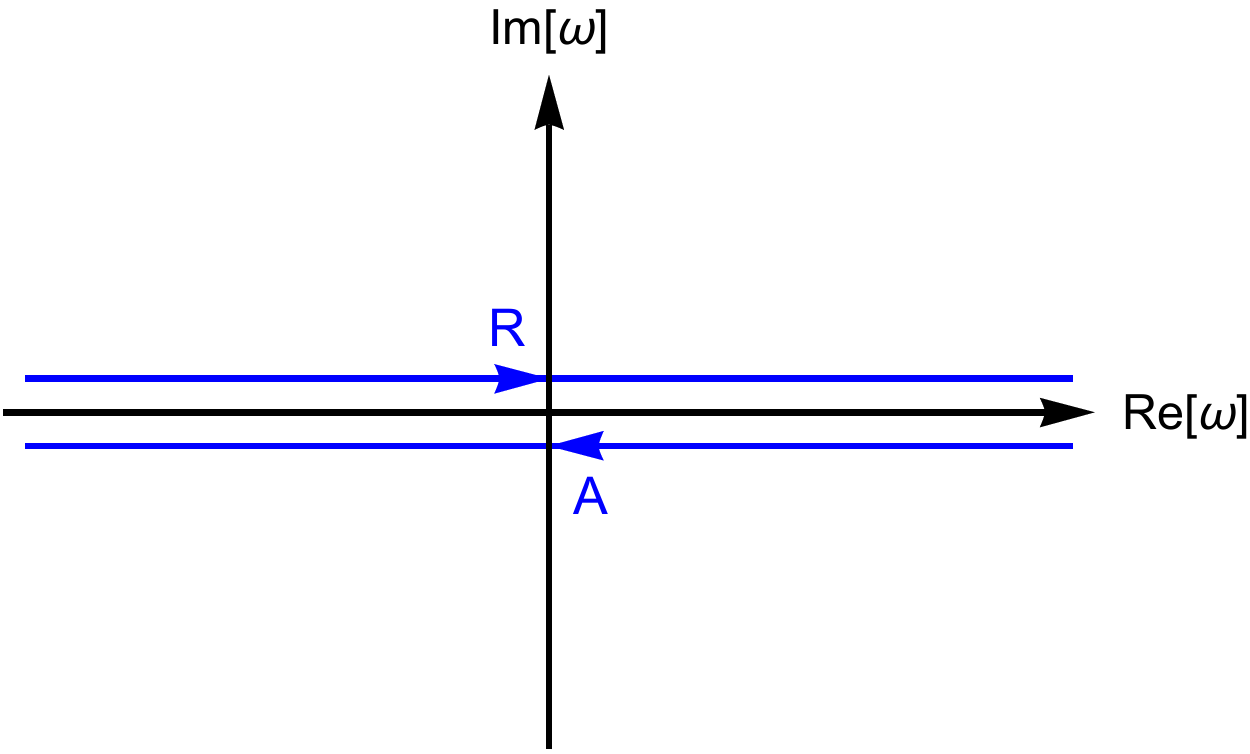}
\caption{The contour for $\oint d\omega$ in Eq.~\eqref{eq:AHESummary}. ``R'' and ``A''
indicate that the Green's functions corresponding to the two branches are retarded and advanced, respectively.  }
\label{fig:contour}
\end{center}
\end{figure}

The central result of the formalism in this paper is a consolidated form
for \textit{all} anomalous transport coefficients
derived in the presence of inelastic dissipative processes. This is given by
\begin{widetext} 
\begin{align} \label{eq:AHESummary}
\mathcal{L}_{ij}^{\alpha \beta}  =  \epsilon_{zij} \epsilon_{z ab } \frac{q^{2-\alpha -\beta}}{2 T^\beta}  \intdk \left\{ \intdw   \Tr \left[  v_a  G_{\R} v_b  G_{\A} \right] \omega^{\alpha+\beta} n_F^{(1)}  + \ointdw  \Tr \left[v_b (\partial_\omega G) v_a G \right]  I^{\alpha \beta} (\omega) \right\},
\end{align}
\end{widetext}
which applies to a general multi-orbital lattice model in two-dimensions. 
For brevity and when it is self evident we have suppressed arguments of the right-hand side variables; we note that
$\epsilon$ refers to the fully antisymmetric tensor
with 
$\{i,j \}=\{x,y\}$. $q=-e<0$ is the charge of electrons. 
The trace $\Tr[\cdots]$ is with respect to the multi-orbital subspace.
$v_{a} \equiv \partial_{k_a} H(\vec{k})$ is the velocity, where
$H(\vect{k})$ is the corresponding multi-orbital Hamiltonian. 
%Both the velocity and Green's functions are matrices in the orbital subspace. 
In the first term inside the curly bracket of Eq.~\eqref{eq:AHESummary},
$n_F^{(1)} \equiv -\partial_\omega n_F(\omega)$ where $n_F(\omega)$ is the Fermi-Dirac distribution function. 
In the second term, the $\oint d\omega$ integral is along the Schwinger-Keldysh contour in Fig.~\ref{fig:contour}. 
The $G$ in $\Tr \left[v_b (\partial_\omega G) v_a G \right] $ can be either $G_{\R}$ or $G_{\A}$, depending on whether $\omega$ is on the forward or backward branch of the contour. 
The functions $I^{\alpha \beta}(\omega)$ are given by 
\begin{subequations} \label{eq:Ialphabeta}
\begin{align} 
I^{00}(\omega)   &  =  n_F(\omega), \\
I^{01}(\omega)  &  = I^{10}(\omega) =   \omega n_F(\omega) + \frac{1}{\beta} \ln\left(1+ e^{-\beta \omega}\right) ,  \label{eq:I01} \\
I^{11}(\omega)  & =  \frac{1}{\beta^2}   \int_{\beta \omega}^{\infty} \frac{x^2}{4 \cosh^2( x/2)} dx ,  \label{eq:Lw}
\end{align}
\end{subequations}
where $\beta=1/T$. 
%The velocity $H(\vect{k}), v_{a}$, and and the self energies $\Sigma_{\R,\A}$ are all matrices in orbital subspace, as is $G_{\R,\A}$.
%In the second term inside the curly bracket of Eq.~\eqref{eq:AHESummary}, the $\oint d\omega$ integral is along the Schwinger-Keldysh contour in Fig.~\ref{fig:contour}. 
%The $G$ in $\Tr \left[v_b (\partial_\omega G) v_a G \right] $ can be either $G_{\R}$ or $G_{\A}$, depending on whether $\omega$ is on the forward or backward branch of the contour. 

When $\Sigma_\R=\Sigma_\A \equiv 0$, the four transport coefficients $\mathcal{L}^{\alpha \beta}_{ij}$ reduce
to known results in the literature. They can be put into the following compact form~\cite{Appendices}
\begin{equation}
\mathcal{L}_{xy}^{\alpha \beta} = \frac{q^{2-\alpha-\beta}}{T^{\beta}} \int d\omega \;   \widetilde{\sigma}_{xy}(\omega) \omega^{\alpha +\beta}  n_F^{(1)}(\omega), 
\label{eq:cleansigmaxy}
\end{equation}
where
\begin{equation}
\widetilde{\sigma}_{xy}(\omega) \equiv \sum_n \int \frac{d\vec{k}}{(2\pi)^2}  \Omega^z_n(\vec{k}) \Theta\left(\omega -\xi_n (\vec{k})\right). 
\end{equation}
Here, $\Omega_n^z(\vec{k})$ is the Berry curvature of the $n$-th energy band of $H(\vec{k})$, with $\xi_n(\vec{k})$
the corresponding energy eigenvalue, and $\Theta(x)$ is the Heaviside function. 
At $T=0$, one can replace $\widetilde{\sigma}_{xy}(\omega)$ with $\widetilde{\sigma}_{xy}(\omega=0)$ in the integrand of Eq.~\eqref{eq:cleansigmaxy}, 
and readily verify that (1): $\sigma_{xy}$  and $\kappa_{xy}$ satisfy the Wiedemann-Franz relation
\begin{equation} \label{eq:WF}
\frac{\kappa_{xy}}{T \sigma_{xy}} = 
L_0 \equiv  \frac{\pi^2}{3} \left( \frac{k_B}{q}\right)^2,
\end{equation}
and (2): $\beta_{xy}$ and $\sigma_{xy}$ satisfy the Mott relation
\begin{equation} \label{eq:Mott}
\beta_{xy} =  \left.\frac{\pi^2}{3} \frac{k_B}{q} k_BT \frac {\partial \widetilde{\sigma}_{xy}(\omega)} {\partial \omega}\right|_{\omega=0}. 
\end{equation}
Here we have restored $k_{B}$. 
These results will not apply except at the lowest temperatures where inelastic dissipation is negligible.

\section{Derivation of anomalous thermal transport coefficients}
\label{sec:Transport}

\subsection{Introduction to the heat-bath approach}
\label{sec:heatbath}

We focus in this section on the thermal Hall coefficient
$\kappa_{xy}$ as given in Eq.~\eqref{eq:AHESummary}. 
This derivation presents a prototype for the other transport coefficients~\cite{Appendices}.

Figure~\ref{fig:Fig1} presents a schematic illustration of our system. It consists of two parts, the 
primary component involves electrons associated with multiple orbitals and confined to a lattice.
The secondary system is a heat bath with an infinite number of localized electronic degrees of freedom.
The corresponding Hamiltonian consists of
three components referring to the particles ($s$), the bath ($b$), and
their coupling ($sb$) such that $ H   = H_s+ H_b + H_{sb}$ with
\begin{subequations} \label{eq:HsHbHsb}
\begin{align}
H_s        & =  \sum_{ij} \sum_{mn} \psi^\dagger_{i m} H_{im, j n} \psi_{jn} - \mu_F  \sum_{im} \psi^\dagger_{im} \psi_{im},  \\
H_b       & =  \sum_{i \alpha} a_\alpha  \phi_{i\alpha}^\dagger \phi_{i\alpha}, \label{eq:Hb} \\
H_{sb}  & =\sum_{i \alpha m} \eta_{\alpha} \psi_{i m}^\dagger \phi_{i \alpha} +\mathrm{h.c.}. 
\label{eq:spectral1}
\end{align}
\end{subequations}
$H_s$ describes a general tight-binding model on a lattice where $\{i,j\}$ represents
lattice sites and $\{m,n\}$ reflects orbital degrees of freedom. 
Here $\psi_{im}$ represents the annihilation operators of the particles of orbital $m$ and at site $i$ in the system,
and $H_{im,jn}$ is the corresponding tunneling between $i$ and $j$. 
Finally, $\mu_F$ is the chemical potential. 

In the heat-bath component described by Hamiltonian $H_b$, $\phi_{i \alpha} $ is the
corresponding annihilation operator at site $i$, while $\alpha$ labels
a characteristic quantum number which, for definiteness, is taken to assume a continuous spectrum~
\footnote{The form of this spectrum is not important for our discussion.}.
We have also assumed that at equilibrium the heat bath is translationally invariant  so that 
the energy level $a_\alpha$ is independent of site index $i$.
The coupling constant $\eta_{\alpha}$ in $H_{sb}$ is similarly spatially independent reflecting translational invariance. 

From the Hamiltonian $H$ one can write down two coupled equations of motion (EOM) for both $\psi_{im}$ and $\phi_{i\alpha}$ fields.
Because of the local nature of the heat-bath particles, one can integrate out the heat bath
and obtain 
\begin{multline}\label{eq:psimotion}
i\evolve \psi_{im}(t) =\sum_{jn} ( H_{im,jn}  -\mu_F \delta_{ij} \delta_{mn} )\psi_{jn}(t)  \\
+D_i(t)  + \int_{-t_0}^{t}\dif t'\Sigma \bigl(t-t') \psi_{im}(t^\prime), 
\end{multline}
where
\begin{subequations}
\begin{align}
D_i(t) &  = \sum_{\alpha} \eta_{\alpha}    \exp\big(-ia_\alpha (t-t_0)\big)  \phi_{i\alpha}(t_0), \label{eq:D} \\
\Sigma(t)  & = \sum_{\alpha}  |\eta_{\alpha}|^2 (- i) e^{ - i a_\alpha t}.  \label{eq:Sigmat}
\end{align}
\end{subequations}
Here, the initial time $t_0$ can be set to $-\infty$ at the end of the procedure. 
Equation~\eqref{eq:psimotion} takes the form associated with generalized Langevin dynamics~\cite{Kubo1966,Greiner1998};
this equation shows that $\psi_{im}$ evolves with time under the influence of the tight-binding Hamiltonian $H_s$, but is
also subject to the ``random force'' $D_i(t)$ with an embedding self-energy kernel $ \Sigma(t-t^\prime)$
representing a ``retarded effect of the frictional force''~\cite{Kubo1966}. 

A local thermal equilibrium of the $\psi_{im}$ field is maintained through a thermal ensemble average of the
heat bath field $\phi_{i\alpha}$ or, equivalently, the source field $D_i$. 
We assume  that $\phi_{i\alpha}$ satisfies ideal Fermi-Dirac statistics
and define 
$\langle \cdots \rangle$ to represent a thermal ensemble average,
so that
$\langle \phi^\dagger_{i\alpha}(t_0)\phi_{j \beta}(t_0) \rangle
=\delta_{ij} \delta_{\alpha\beta} n_F(a_{\alpha})$ at time $t_0$.
We then find
\begin{multline}\label{eq:DDexpect}
\langle D_{i}^\dagger(t)D_{j}(t^\prime) \rangle= \delta_{ij}  \sum_{\alpha} |\eta_\alpha|^2 e^{i a_\alpha(t-t^\prime)}  n_F(a_\alpha). 
\end{multline}
Fourier transformation of this relation in frequency and $\vec{k}$ space leads to~
\footnote{In our convention, the time-frequency and space-momentum Fourier transformations are defined as $f(t)= \intdw f(\omega) e^{-i \omega t}$
and $f(\vec{x})= \intdk f(\vec{k}) e^{+i \vec{k}\cdot \vec{x}}$, respectively. For brevity, we also use the same notation for functions before and after
the transformations. }
\begin{align} \label{eq:FDT}
\langle D^\dagger D \rangle(\omega) =  - 2 \ImSigma(\omega)  n_F(\omega),
\end{align}
which importantly represents a generalized fluctuation-dissipation theorem (FDT). 
The left-hand side characterizes fluctuations of the bath degrees of freedom while the right-hand side depends on
the imaginary part of the retarded self energy $\Sigma_{\R}$ of the particles in the system and
thus corresponds to the dissipation they experience. 
Equation~\eqref{eq:FDT} will play an important role in our derivation of transport coefficients in Sec.~\ref{sec:kappaxyderivation}. 

Here $\Sigma_{\R} \equiv \ReSigma + i \ImSigma$ is the Fourier transform of $\Theta(t)\Sigma(t)$.
From Eq.~\eqref{eq:Sigmat} we have
\begin{subequations}
\begin{align}
\ImSigma(\omega) & \equiv - \pi \sum_{\alpha} |\eta_{\alpha}|^2 \delta(\omega - a_\alpha), \\
\ReSigma(\omega) & \equiv \mathcal{P} \int \frac{d\omega^\prime}{\pi} \frac{  \ImSigma(\omega^\prime)}{\omega^\prime -\omega}, \label{eq:KK}
\end{align}
\end{subequations}
where $\mathcal{P}$ represents the Cauchy principal value. 

From the Fourier transform of Eq.~\eqref{eq:psimotion}, one can derive
\begin{align}
\Glesser_{m,n}(\vec{k},\omega)  &  \equiv i  \langle \psi_{m}^\dagger(\vec{k},\omega) \psi_{n}(\vec{k},\omega) \rangle  \nonumber \\
& = i  \big[ G_{\R}(\vec{k},\omega)\langle D^\dagger D\rangle(\omega) G_{A}(\vec{k},\omega)\big]_{m,n},
\label{eq:Glesser0}
\end{align}
where $\Glesser$ is the conventional ``lesser'' Green's function~\cite{Mahan2000}. 
Just like $G_\R$ and $G_\A$, $\Glesser$ is a matrix in the orbital subspace.
Because the heat-bath degrees of freedom are orbitally independent,
both $\langle D^\dagger D\rangle(\omega)$
and $\Sigma(\omega)$ are local functions proportional to the identity matrix in this space.

In the presence of external perturbations, $\vec{E}$ or $ \vect{\Theta}=- \nabla T$, $\Glesser$
deviates from its equilibrium value in Eq.~\eqref{eq:Glesser0}, and the deviation
can be computed to linear order in $\vec{E}$ or $ \vect{\Theta}$ using standard perturbation theory. 
From the perturbed $\Glesser$ one can then compute the ``microscopic'' charge
and heat-current densities, $\langle \hat{ \vect{J}}^e \rangle_{\vect{E}, \vect{\Theta}} $ and $\langle \hat{\vect{J}}^h \rangle_{\vect{E}, \vect{\Theta}}$,
from which the transport coefficients can be derived after appropriate magnetization current subtractions.

\subsection{Evaluation of $\kappa_{xy}$} 
\label{sec:kappaxyderivation}

As a prototypical example of the subsequent implications for transport, we evaluate the
thermal Hall conductivity, which describes a transport heat-current response to the perturbation $ \vect{\Theta}$. 
The heat-current density operator satisfies the continuity equation~\cite{Appendices}.
Its expectation value can be written in coordinate and Fourier transform space as
\begin{widetext}
\begin{align} \label{eq:JhgradT}
\langle \hat{\vec{J}}^h \rangle_{\gradT} & = \bigg \langle  \frac{1}{V} \int d \vecx dt  \int d \vecxp d t^\prime  
 \mathrm{Tr} \bigg[ \psi^\dagger(\xptp ) \vec{v}(\vecxp, \vecx) \frac{ i \overrightarrow{\partial_t }- i \overleftarrow{ \partial_{t^\prime}} }{2} \psi(\xt)  \bigg]  \bigg\rangle 
 = \frac{1}{i}\int \dkdw  \mathrm{Tr} \left[ \vec{v}(\vec{k})  \omega \Glesser_{\gradT}(\vec{k},\omega)\right], 
\end{align}
\end{widetext}
where, for simplicity, we have switched from a discrete lattice sum $\sum_i$ to a continuous $\int d\vec{x}$ integral notation~\cite{Appendices},
and dropped the orbital index of all operators. 
$V$ is the spatial volume; 
$\vec{v}$ is the velocity operator defined from $H_s$. 
The subscript $\gradT$ in $\Glesser_{\gradT}$
indicates that this is the perturbed Green's function, whose derivation closely follows that of Eq.~\eqref{eq:Glesser0}.

In the presence of $\gradT$, the only difference is that in the equation of motion in Eq.~\eqref{eq:psimotion}, we now have $D_i \rightarrow D_i -  \frac{\partial D_i}{\partial T} \vec{x}\cdot \gradT$
and $\Sigma \rightarrow \Sigma -  \frac{\partial \Sigma}{\partial T} \vec{x} \cdot\gradT$ to linear order in $\gradT$~
\footnote{This is because in the presence of a temperature gradient, in order to maintain a local equilibrium, the heat bath particle energy levels $a_\alpha$ needs to depend on the local
temperature $T(\vec{x})$ at $\vec{x}$, leading to $a_\alpha \rightarrow a_\alpha - ( \partial a_\alpha/\partial T) \vec{x}\cdot \gradT$ in a linear expansion,
which in turn leads to a shift in $D_i$ and $\Sigma$.}.  Making these substitutions in Eq.~\eqref{eq:Glesser0} leads to
\begin{widetext}
\begin{equation} \label{eq:GlesserGradT}
\Glesser \rightarrow \Glesser_{\gradT} =  \Glesser -  i \left[ G_\R  \frac{ \partial\Sigma_\R } { \partial T} \vec{x} G_\R \langle D^\dagger D \rangle G_\A
+ G_\R \langle D^\dagger D \rangle G_\A    \frac{ \partial\Sigma_\A } { \partial T}  \vec{x} G_\A 
+ G_\R \left( \left \langle \frac{\partial D^\dagger}{\partial T}  D \right\rangle 
+ \left \langle D^\dagger \frac{\partial D}{\partial T} \right \rangle \right) \vec{x}  G_\A  \right] \cdot \gradT,
\end{equation}
\end{widetext}
where we have suppressed the arguments $\vec{k}$ and $\omega$ of all functions~
\footnote{Note that, on the right-hand side, $\vec{x}$ is an operator which is nonlocal in $\vec{k}$ space and it is, therefore, important
to keep track of its order with respect to other functions inside the expression. 
In Eq.~\eqref{eq:GlesserGradT}, $\vec{x}$ operates on all functions to its right.}. 
Substituting the result for $\Glesser_{\gradT}$ into Eq.~\eqref{eq:JhgradT} and replacing $\vec{x}$ with $i \nabla_{\vec{k}}$~
\footnote{We note that although defining the position operator $\vec{x}$ properly for a periodic lattice is subtle~\cite{Blount1962,Resta2018,Ventura2017,Parker2019}, 
within a gradient expansion due to a slowly varying external perturbation it is simply $i \nabla_{\vec{k}}$ in the $\vec{k}$ space. 
The subtlety does not come in here because the gradient expansion is a local one which does not depend on how one chooses the origin for $\vec{x}$ in spatial coordinates. }
leads to $\langle  J_i^h \rangle_{\gradT} =
L^{h,h}_{ ij} \Theta_j$ with
\begin{widetext}
\begin{align} \label{eq:Lhh}
L^{h,h}_{ ij} =  &  -  \frac{\epsilon_{zij} \epsilon_{zab}}{2} \intdkdw \Tr \left\{ v_a \omega 
\left[ 
  G_\R    \frac{\partial \Sigma_\R }{\partial T} \partial_{b} \Glesser
 + \Glesser \frac{\partial \Sigma_\A}{\partial T} \partial_{b}  G_\A 
  +  i G_\R    \left(  \frac{\partial}{\partial T} \langle  D^\dagger D \rangle \right)   \partial_{b} G_\A
 \right] \right\} , 
\end{align}
\end{widetext}
where $\partial_b \equiv \partial_{k_b}$. The $\epsilon_{zij} \epsilon_{zab}/2$ factor is included since only the $i \leftrightarrow j$ antisymmetric part contributes
to anomalous thermal Hall transport. 

As discussed in Sec.~\ref{sec:general}, $\langle \hat{\vec{J}}^h \rangle_{\gradT} $ is not the transport current. To obtain the latter, one needs to subtract the heat magnetization current, 
$\partial \vec{M}^h/\partial T \times \gradT$, from $\langle \hat{\vec{J}}^h \rangle_{\gradT} $. 
An expression for $\vec{M}^h$ can be derived~\cite{Appendices} by considering a semi-infinite system with an edge and assuming that the 
chemical potential $\mu_F$ varies slowly from its bulk value far away from the edge to $\mu_F=-\infty$ (i.~e., its corresponding vacuum value) just right outside of the edge.
Because of spontaneous time-reversal-symmetry breaking due to orbital motion, there is a heat current (as well as an electric current) flowing along the edge, whose density $\vec{J}^h_{\mathrm{edge}}$
can be related to the bulk heat magnetization by $\vec{J}^h_{\mathrm{edge}}=\nabla \mu_F \times \frac{\partial \vec{M}^h}{\partial \mu_F} $, where $\nabla \mu_F$ is the spatial gradient of $\mu_F$ due to the
edge.  $\vec{J}^h_{\mathrm{edge}}$ can be evaluated just as $\langle \vec{J}^h \rangle_{\gradT}$ by including $(1-\partial \Sigma/\partial \mu_F)\vec{x}\cdot \nabla \mu_F$ 
as a perturbation to the EOM in (the time-frequency Fourier transformed) Eq.~\eqref{eq:psimotion}. 
From $\vec{J}^h_{\mathrm{edge}}$ one then obtains $\partial \vec{M}^h / \partial \mu_F$, and integrating over $\mu_F$ yields
\begin{subequations} \label{eq:Mzh}
\begin{align} 
 M_z^h   & =   \frac{\epsilon_{z ij}}{2} \intdkodw  \mathrm{Tr} [ v_i  (\partial_{\omega} G) v_j G]  Q(\omega),  \\
 \rm{with} \quad Q(\omega)  &   \equiv \frac{\omega}{\beta} \Li_1 \left(- e^{-\beta \omega}\right)   + \frac{1}{\beta^2}\Li_2\left(- e^{-\beta \omega}\right),
 \label{eq:Qdef}
 \end{align}
\end{subequations}
where $\oint d\omega$ is along the contour in Fig.~\ref{fig:contour}. 
Because we are considering two dimensions, only the $z$-component of $\vec{M}^h$ is relevant. 
In the equation above, $ \Li_n(x)$ is the $n$-th polylogarithm function, which is defined in general for arbitrary complex order $s$ and $|z|<1$ by 
\begin{equation}
\Li_s(z) =  \sum_{k=1}^{\infty} \frac{z^k}{k^s}. 
\end{equation}
For $s=1$, $\Li_1(z)=-\ln(1-z)$. 

From Eq.~\eqref{eq:Currents} we obtain
\begin{align} \label{eq:kappaxysplit}
\kappa_{ij} & =  L_{ij}^{h,h} - \epsilon_{i z j} \partial M_z^h/\partial T \equiv \kappa_{ij}^{(1)} +\kappa_{ij}^{(2)}, 
\end{align}
where $\kappa_{ij}^{(1)}$ contains all terms that involve $\partial \Sigma/\partial T$ from both $L_{ij}^{h,h}$ and $M_z^h$, while $\kappa_{ij}^{(2)}$ contains the remainder.

\begin{figure*}[htp]
\begin{center}
\includegraphics[width=0.8\textwidth, trim=0mm 30mm 0mm 30mm]{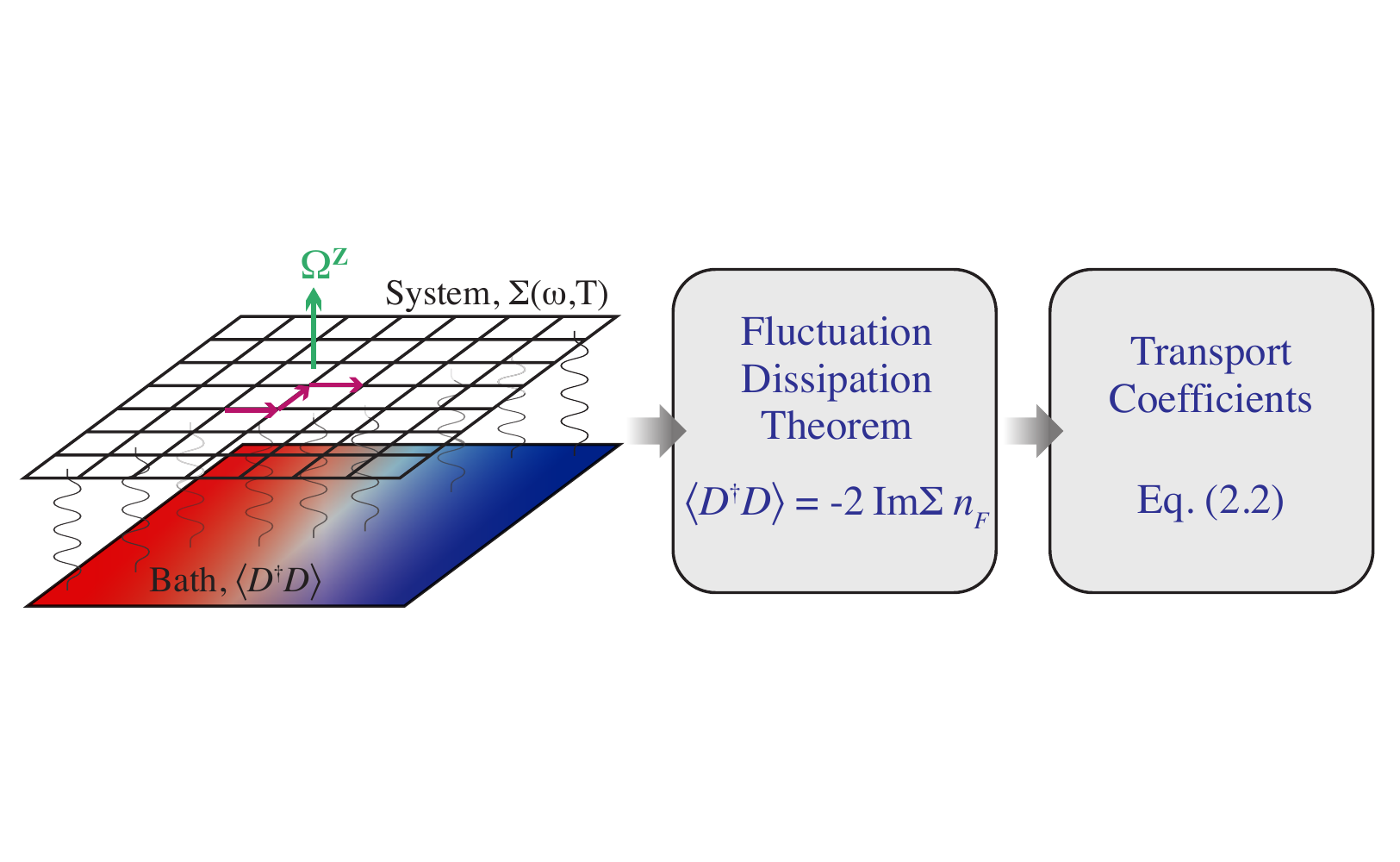}
\caption{Schematic figure emphasizing that because the system and heat bath are coupled and in local equilibrium with each other, the electron self energy $\Sigma$ originating from the coupling and the heat-bath correlator $\langle D^\dagger D\rangle$
are not independent. Instead, they satisfy the fundamental fluctuation-dissipation relation, expressed in Eq.~\eqref{eq:FDT}, which underpins the notion of local equilibrium. This leads to the simple
and rather elegant form of all our derived transport coefficients seen in Eq.~\eqref{eq:AHESummary}.}
\label{fig:illustration}
\end{center}
\end{figure*}

In the following, we show that $\kappa_{ij}^{(1)} =0 $ because of a cancellation due to the fluctuation-dissipation relation in Eq.~\eqref{eq:FDT}. 
Collecting all $\partial \Sigma/\partial T$ terms from Eq.~\eqref{eq:kappaxysplit}, and using Eqs.~\eqref{eq:Lhh}, \eqref{eq:Mzh}, and \eqref{eq:FDT}, we find
\begin{widetext}
\begin{subequations} \label{eq:kappaxy1}
\begin{align}
\kappa_{ij}^{(1)}   &  =   -   \frac{\epsilon_{zij} \epsilon_{zab}}{2}  \intdkdw
 \Tr \left [   \omega v_a  G_\R \frac{\partial \Sigma_{\R} }{\partial T} \partial_b \Glesser 
 + \omega v_a  \Glesser \frac{ \partial \Sigma_{\A} }{\partial T}  \partial_b G_{\A}
 +   \omega v_a   G_{\R} \frac{ \partial (-2 i \ImSigma)}{\partial T} \partial_b G_{\A}   n_F \right] \nonumber \\
 &\quad +  \frac{\epsilon_{zij} \epsilon_{zab}}{2}  \intdkodw  \Tr \left[  v_a     (\partial_T  \partial_\omega  G)  v_b G  \right] Q
\label{eq:kappaxyderivation2} 
\\ &  =   \frac{\epsilon_{zij} \epsilon_{zab}}{2}  \intdkodw  \left\{   \Tr \left[ v_a  (\partial_T G)  v_b G  \right]   \omega n_F
  + \left( \partial_\omega \Tr \left[  v_a   (\partial_T   G)  v_b G  \right] \right)   Q   \right\}
\label{eq:kappaxyderivation3}  \\
 & =0. \label{eq:kappaxyderivation4} 
\end{align}
\end{subequations}
\end{widetext}
The first line comes from $L^{h,h}_{ij}$, while the second arises from $- \epsilon_{i z j} \partial M_z^h/\partial T$.
In the derivation we have used $\Glesser= (G_\A -G_\R)  n_F$.
To arrive at the last line we have performed an integration by parts over $\omega$ in the second term in Eq.~\eqref{eq:kappaxyderivation3} and used the identity
$ \partial_\omega Q =  \omega n_F$. 
{\textit{We emphasize that  the fluctuation-dissipation
relation, Eq.~\eqref{eq:FDT}, is central for the complete cancellation of the $\partial \Sigma/\partial T$ contributions to $\kappa_{ij}^{(1)}$.}}
This is schematically shown in Fig.~\ref{fig:illustration}. 
A similar cancellation occurs in the derivation of the coefficient $\beta_{xy}$~\cite{Appendices}.

What remains in Eq.~\eqref{eq:kappaxysplit} is then 
\begin{widetext}
\begin{align}
\kappa_{ij}^{(2)}   & =  \frac{\epsilon_{zij} \epsilon_{zab}}{2} \intdkdw  \left\{   \Tr [\omega v_a G_\R (  2 i  \, \ImSigma ) G_\A v_b G_\A  ] \; \partial_T n_F 
 +   \ointdw \Tr \left[v_a ( \partial_\omega G) v_b G \right]  \; \partial_T Q \right\}.
\end{align}
\end{widetext}
The first term arises from $L^{h,h}_{ij}$ while the second from $-\epsilon_{izj} \partial M^h_z/\partial T$. 
Using $G_\R^{-1} -G_\A^{-1}= - i 2 \ImSigma$ we find that $\kappa_{ij}^{(2)}$ yields the full
$\kappa_{xy}$ of Sec.~\ref{sec:summary}, where we make use of
$\partial_T n_F=  (\omega/T) n_F^{(1)}$ and $\partial_T Q = - I^{11}(\omega)/T$. 
This completes our derivation of $\kappa_{xy}$.

\section{Numerical Implications}
\label{sec:numerics}

We now explore the numerical implications of our transport coefficient formula, Eq.~\eqref{eq:AHESummary}.
There have been experimental studies of all anomalous coefficients 
~\cite{Shiomi2010, Li2017,Xu2020}, with recent emphasis on 
two important ratios: the first involves a thermoelectric coefficient $\beta_{xy}/\sigma_{xy}$ and the 
second corresponds to the widely studied Wiedemann-Franz ratio $\kappa_{xy}/( \sigma_{xy} T)$.
Importantly, these experiments address the two ratios over the entire
temperature range up to and beyond room temperature, where inelastic scattering effects, 
as in electron-phonon scattering, are expected to play an important role. 

While little is known theoretically about how inelastic scattering
affects the behavior of the anomalous transport coefficients and their ratios, these effects are 
necessarily important once the system has departed from the ground state. 
It is widely believed~\cite{Li2017,Xu2022} that inelastic scattering leads to a depression
rather than an enhancement in the Wiedemann-Franz ratio, once
temperature is increased from zero. This assertion is correct only for
the longitudinal Wiedemann-Franz ratio~\cite{Appendices}; here we show the situation is different for the anomalous transverse case
where Berry-curvature physics becomes important and non-universal behavior obtains.

Using Eq.~\eqref{eq:AHESummary}, we explore the transport properties
for two different band structures showing how $\kappa_{xy}/( \sigma_{xy} T)$
can exhibit complex behavior as a result of the distribution of Berry curvatures. 
We also address the universal behavior of another important  ratio $\beta_{xy}/\sigma_{xy}$.

We consider a generic two-band tight-binding model Hamiltonian which supports non-zero
Berry curvature effects and is of the general form in $\vec{k}$ space
 \begin{gather} \label{eq:modelHk}
H(\vec{k}) = h_0(\vec{k}) +\vec{h}(\vec{k}) \cdot \boldsymbol{\sigma} -\mu_{\mathrm{F}},
\end{gather}
where $\boldsymbol{\sigma} =(\sigma_x, \sigma_y,\sigma_z)$ are the three Pauli matrices defined in $2\times 2$ orbital space.
In the following we assume only one spin component of electrons is populated with a fully polarized magnetic metal in mind. 
However, we expect that our analysis is applicable to a large class of
systems, including for example noncollinear antiferromagnets which have received significant attention~\cite{Xu2020}.
Here we address two related variants of this Hamiltonian~\cite{Appendices}. 
One is topological ($\HII$) while the other ($\HI$) is not.
The topological case derives from a model of a Chern insulator~\cite{Neupert2011} 
with complex hopping integrals associated with
a staggered magnetic flux through each square lattice plaquette.
$\HI$ describes a system of two quasi-1d d-orbitals 
also on a square lattice with a spin-orbit coupling (SOC) that gives rise to nonzero Berry curvature. 
For simplicity we defer the detailed parameterizations of both $\HII$ and $\HI$ to the Appendix~\cite{Appendices}.

The inelastic contributions to $\Sigma(\omega,T)$
are assumed to arise from electron-phonon coupling~\footnote{Because electrons and phonons have different 
statistics and charge, 
this would require a modified treatment of our heat bath.  
However, to the extent that the dissipation we consider enters transport only through a local (i.~e., $\vec{k}$ independent)
electronic self energy, we believe our formulas do not depend on the exact microscopic basis of $\Sigma$.}, which should become relevant at any non-zero temperature.
Its effect on longitudinal transport is relatively well understood~\cite{Lavasani2019,Appendices}, and so serves as a useful baseline. 

Quite generally one models the electron-phonon scattering to be associated with an electron self energy~\cite{Mahan2000}, 
\begin{widetext}
\begin{align}  \label{eq:ImSigmaDef}
\mathrm{Im}\Sigma(\omega) & =  -\pi \int_0^{\omega_D} d\omega^\prime   \alpha^2 F(\omega^\prime) [ 2 n_B(\omega^\prime)  + n_F(\omega^\prime + \omega) + n_F(\omega^\prime - \omega) ],  
\end{align}
\end{widetext}
where $n_B(\omega)=1/(e^{\omega/T }-1)$. Here,  $\alpha^2 F(\omega)$ reflects the coupling~\cite{Mahan2000} 
and $\omega_D$ is the Debye frequency beyond which $\alpha^2 F(\omega) $ vanishes.
For acoustic phonons, which are dominant, the related spectral function can be modeled by~\cite{LaShell2000}
\begin{equation}
\alpha^2 F(\omega)=\lambda (\omega/\omega_D)^2 \Theta(\omega_D - |\omega|). 
\end{equation}
$\ReSigma$ is then obtained by using the Kramers-Kronig relation, Eq.~\eqref{eq:KK}.

\begin{figure*}[htp]
\includegraphics[width=\textwidth,trim=0mm 0mm 0mm 0mm]{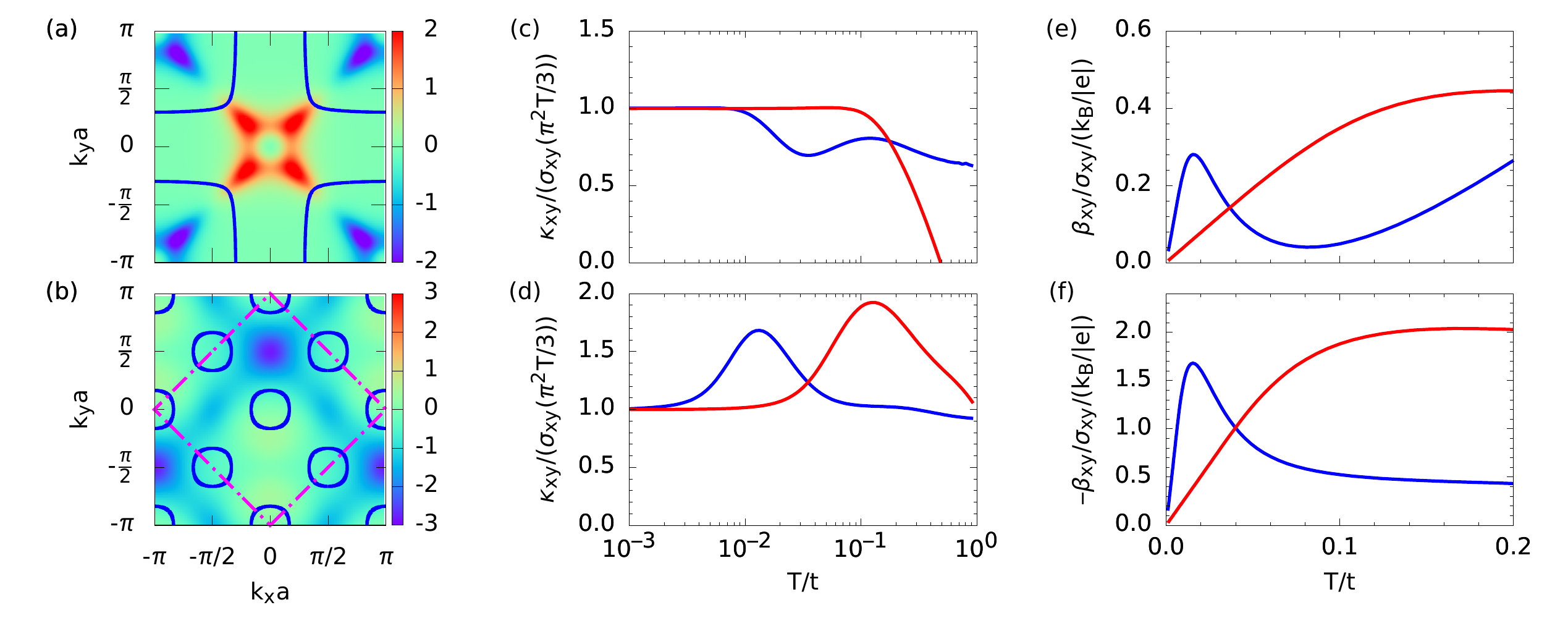}
\caption{The effects of two different Berry-curvature distributions, (a) and (b), on anomalous transport coefficient ratios, (c)-(f), 
showing the effects with (blue) and without (red) dissipation.
The Berry curvature is plotted for the lower energy band together with the corresponding Fermi surface contour (blue solid line). 
The magenta dashed line in (b) denotes the reduced Brillouin zone.
Upper panels (a), (c) and (e) are for model $\HI$ and lower panels (b), (d) and (f) are for model $\HII$,
whose energy bands are topologically non-trivial.
Note the additional minus sign in the vertical axis label in (f).
For the simulations in (c-f), we have chosen the Debye frequency $\omega_D=0.1 t$, where $t$ is the nearest-neighbor hopping integral in each model
and the electron-acoustic-phonon coupling strength $\lambda=5$. A large $\lambda$ is chosen for a better visualization of inelastic scattering effects.}
\label{fig:WFBerry}
\end{figure*}

Figure~\ref{fig:WFBerry} 
addresses the behavior of the anomalous Wiedemann-Franz ratio $\kappa_{xy}/(\sigma_{xy} T)$ for both models.
The Berry-curvature distribution for the two models is plotted
in  Figs.~\ref{fig:WFBerry} (a) and (b).
Figures~\ref{fig:WFBerry} (c) and (d) present plots comparing
the behavior with inelastic scattering included (blue) and without (red). 

If for the moment we omit inelastic effects there are some interesting observations to be made.
It has been claimed in the literature~\cite{Xu2020} that, when
$T$ increases, $\sigma_{xy}$ and $\kappa_{xy}$ sample the same Berry-curvature distribution, 
but with different weights for a given energy due to the $\omega^{\alpha+\beta}$ factor in Eq.~\eqref{eq:cleansigmaxy}.
This difference in weighting, combined with the dispersion of Berry-curvature distribution with energy, 
leads to different temperature dependences of $\sigma_{xy}$ and $\kappa_{xy}/T$
and, therefore, to a violation of the WF law once temperature becomes sufficiently high.
The direction of the violation of the WF ratio has been argued \cite{Xu2020} to be consistent: one always
sees a downward deviation of  $\kappa_{xy}/(\sigma_{xy} T)$ from the Lorenz number at finite temperature~\cite{Xu2020}.

As can be seen from the red curves in the plots, what we observe (even without dissipation) is rather different. 
We find the WF ratio can be either enhanced  (red curve in Fig.~\ref{fig:WFBerry}(d))
or suppressed  (red curve in Fig.~\ref{fig:WFBerry}(c))
relative to the $T=0$ value. 
These differences reflect the fact that
Berry curvatures from different regions of reciprocal space carry opposite signs and thus compete.
Thus, the overall behavior of the WF ratio reflects both
the sign competition of Berry curvatures as well as the different frequency weightings in heat and charge transport.
Finally, we emphasize that having a topologically nontrivial band is not a necessary condition for an enhanced WF
ratio, as shown in Fig.~\ref{fig:WFBerry} (d), red curve,
as we have seen a similar enhancement in some trivial band models as well.
 
Once inelastic scattering is included the effects of
non-zero $\Sigma(\omega,T)$ add to the complications discussed above.
With this temperature-dependent dissipation, the association with Berry curvature 
deriving from Eq.~\eqref{eq:AHESummary}
is no longer given by Eq.~\eqref{eq:cleansigmaxy}. 
Nevertheless one can still use the latter equation and the concept of Berry curvature as a guideline, and infer that the presence of $\Sigma(\omega, T)$ will change the 
Berry-curvature contributions to each transport coefficient and
lead to either an additional upturn deviation or a downturn deviation from the ``baseline'' WF ratio. 
This will  be apparent only within the inelastic regime ($T \lesssim \omega_D$). 

These violations due to the interplay between the Berry-curvature distribution and inelastic scattering are seen in the two
models and plotted as the blue curves in Figs.~\ref{fig:WFBerry} (c) and (d).
For completeness we note that finite-temperature enhancements of the WF ratio have been 
recently observed in a kagome Chern magnet~\cite{Xu2022}.
Overall, our findings highlight the intrinsic challenge of disentangling inelastic scattering from Berry-curvature 
effects in analyzing experimental data. Moreover, they point out that unlike the longitudinal case
the behavior of the temperature-dependent transverse WF ratio is expected to be highly non-universal.

\begin{figure}[htp]
\includegraphics[width=\linewidth,trim=0mm 0mm 0mm 0mm]{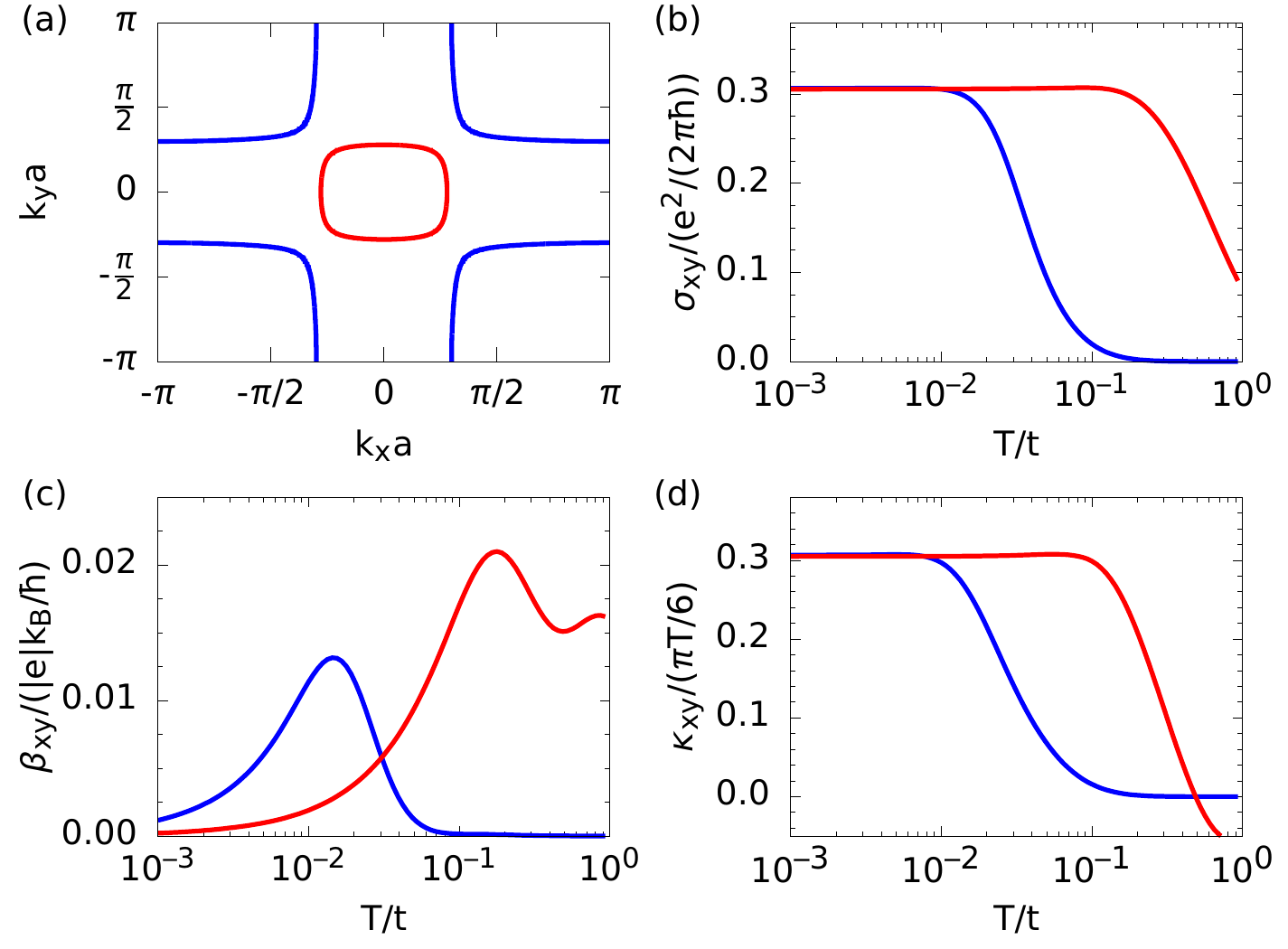} 
\caption{Fermi surface (a) and complete set of anomalous transport coefficients (b-d) for the non-topological model, $\HI$.
Similar results for $\HII$ are presented in the Appendix.
In (b-d), blue and red lines correspond to results with and without inelastic scattering effects, respectively.
Note that the transport coefficient $\gamma_{xy}$ is not independent (due to the Onsager relation) and
not plotted here.}
\label{fig:KontaniTransport}
\end{figure}

Although we have focused our discussion on the ratio $\kappa_{xy}/(\sigma_{xy} T)$, 
we  point out that this
sign competition (and entanglement of inelastic dissipation with Berry curvature) is reflected in
all anomalous transport, not just in the ratios but also in
$\sigma_{xy}$ and $\kappa_{xy}/T$, as shown by
the red and blue curves in Figs.~\ref{fig:KontaniTransport} (b) and (d). 
We also emphasize that
there is a similar interplay between Berry curvature and inelastic self-energy effects in the
anomalous thermoelectric parameter, $\beta_{xy}$, as shown in Fig.~\ref{fig:KontaniTransport} (c).
However, interestingly, we find that, in contrast to $\kappa_{xy}/(\sigma_{xy} T)$, the ratio $\beta_{xy}/ \sigma_{xy}$ seems
to exhibit a universal behavior: we consistently find that the low-temperature slope of $|\beta_{xy}/ \sigma_{xy}|$ 
appears to be enhanced by inelastic scattering.
This is shown for the model $H_{\mathrm{I}}$ in Fig.~\ref{fig:WFBerry}(e), while
Fig.~\ref{fig:WFBerry}(f) plots the corresponding result for the topological case of $H_{\mathrm{II}}$. 
For the latter, the same enhancement appears despite the fact that $\beta_{xy}/\sigma_{xy}$ now has the opposite sign~
\footnote{We have also checked that this enhancement occurs in all the models we have studied, including different variants of $H_{\mathrm{II}}$, and we believe that it is universal.} . 
We can speculate, moreover, that our results may provide an explanation for the recently observed enhancement of 
$|\beta_{xy}/\sigma_{xy}|$ at $T\lesssim 200 K$ in the kagome Chern magnet TbMn$_6$Sn$_6$~\cite{Xu2022}, 
although a much more detailed study is needed to confirm this.

We believe this universal enhancement of the ratio $\beta_{xy}/\sigma_{xy}$
can be roughly understood as follows. 
We first notice that, in the absence of inelastic scattering,
the ratio can be interpreted as
a ratio between the transverse flow of thermal entropy and that of charge, both
under the influence of Berry curvature~\cite{Li2017,Ding2019,Xu2020},
averaged over different energy eigenstates.  
Note that $\sigma_{xy}$ can be written as a sum of Berry curvature from \textit{all filled} states as incorporated in the occupation number $n_F$.
In contrast, $\beta_{xy}$ is a sum of the same Berry curvature but now convoluted with the thermal entropy -- $I^{01}$
in Eq.~\eqref{eq:I01} -- which is nonzero only for \textit{partially occupied states} that lie within an energy shell $\sim T$ near the Fermi level.

Now, if we include the self energy $\Sigma$, $\sigma_{xy}$ and $\beta_{xy}$ respond quite differently due to the key difference between the two thermal factors, $n_F$ and $I^{01}$. 
At low temperature, $\Sigma$ is small and its effect is significant only on states within an energy window $\sim |\Sigma|$ near the Fermi level, not those deep in the Fermi sea. 
Consequently, $\sigma_{xy}$ is roughly unchanged (see curves at $T/t \lesssim 0.01$ in Fig.~\ref{fig:KontaniTransport}(b)), 
while $\beta_{xy}$ is much more strongly enhanced by the presence of $\Sigma$ (see Fig.~\ref{fig:KontaniTransport}(c)).
This latter enhancement comes from an energy broadening due to $\ImSigma$, which has a similar effect on thermal entropy as an increase of temperature.  
In summary, the net effect of $\Sigma$ at low temperature
is an overall enhancement of $\beta_{xy}$ and of $|\beta_{xy}/\sigma_{xy}|$ as well.

\section{Conclusions and Outlook}
\label{sec:conclusion}

This paper provides a basis for understanding what happens to anomalous transport properties when
temperature is increased beyond the lowest temperature regime. We consider this important question,
relevant to a wide class of experiments (as phonons are inevitably present), in the context
of anomalous thermal Hall and other anomalous thermoelectric transport coefficients. 
Our methodology is new and addresses an
interesting paradox posed when thinking about how a temperature-dependent self energy can be consistently
addressed in the presence of a thermal gradient. 
Here we emphasize the advantages of
a bath approach as distinct from Luttinger's somewhat controversial scheme
~\cite{Kapustin2020,Stone2012,Tatara2015,Gromov2015,Shitade2014,Park2021}.

Part of the failure to address finite temperature and attendant inelastic scattering
in past theoretical literature is likely based on the presumption that these effects in
the transverse channel are not that different from those in the longitudinal case.
On this basis one might expect that inelasticity
universally suppresses the Wiedemann-Franz ratio.

We show here this reasoning does not hold and that the effects of inelasticity in the transverse
channel are strongly intertwined
with the detailed distribution of the Berry curvature. 
Importantly, there are few universal features one can predict in advance about
the temperature-dependent behavior of the transverse Wiedemann-Franz ratio. 
By contrast, we are able to identify a transport ratio involving a thermoelectric
response, $\beta_{xy}/\sigma_{xy}$,  which exhibits significant universality
in the low-temperature regime; importantly,
this occurs even in the presence of Berry curvature and inelastic dissipation.
While there is some experimental support for both these
observations~ \cite{Xu2022}, it is hoped that these effects and
the expected universality can be explored in more detail in future literature.

Our heat-bath approach leads to a
consolidated formula for the entire class of transport coefficients.
Indeed, bath approaches have been advocated rather widely within different
subfields of the physics community,
particularly for electric-field-driven transport. We argue that such
approaches are most ideally suited to situations when
a temperature gradient or heat current is present, which is not where the
prior widespread interest has been. We show here our bath approach provides
a more intuitive route to thermal transport and also
reveals the fundamental role played by the fluctuation-dissipation theorem.

Also notable is that
our heat-bath approach is readily adapted to address
transverse thermal transport in a small magnetic field in the presence of
inelastic processes; this is associated with conventional
thermoelectric and thermal Hall transport.
Here, as in the present paper and for
a generic multi-band lattice model~\cite{Nourafkan2018}, 
one can derive a consolidated set of transport coefficients. 

For concreteness we have emphasized inelastic dissipation which originates from one particular
mechanism, namely, electron-phonon scattering.
Nevertheless, we expect our consolidated formulas for all thermal and charge transport coefficients
should be applicable to alternative models with different inelastic processes
underlying the self energy.

Thus we are led to speculate that, while our calculations are exact in the absence of inter-particle
interactions, they may also describe inter-particle interaction effects provided these are associated with a local self energy.
In this way, we subscribe to the philosophy in the literature surrounding the
Kadanoff-Baym approach to nonequilibrium transport which involves interacting particles.
Indeed, the Kadanoff-Baym equation has been re-interpreted as a generalized Langevin equation with non-interacting particles coupled to a heat bath~\cite{Greiner1998},
where the bath is assumed to give rise to self-energy contributions deriving from many-body effects.

As we look to the future, it will be interesting to investigate whether our results
can be applied to study systems having a local self energy 
of the form which appears in studies of quantum criticality.
Here one encounters rather violent quantum fluctuation effects which make it problematic to
address anomalous thermal transport.
Additional new directions involve a generalization to three-dimensional topological
materials such as Weyl semimetals~\cite{Sharma2016}. 
More challenging, finally, is to consider the effects of a
$\vec{k}$-dependent self energy originating from the heat bath. 
Including these non-local effects will be particularly complicated by the fact that the heat bath degrees of freedom can also carry currents if the particles are not localized.
A proper treatment of both the system and heat bath on an equal footing is needed to preserve the charge and energy conservation laws. These are crucial for 
ensuring that, (in the language of Kubo-like diagrammatic approaches), vertex-correction contributions to transport coefficients are properly included.

\section{Acknowledgment}
We thank Michael Levin for insightful discussions and Qijin Chen for past collaboration on related subjects. 
We are grateful to Kamran Behnia and Shuang Jia for their comments on the paper. 
This work was primarily funded by the University of Chicago Materials Research Science
and Engineering Center through the National Science Foundation under
Grant No. DMR-1420709.
It is completed in part with resources
provided by the University of Chicago's Research Computing Center. 
R.B. was supported by D\'epartement de physique, Universit\'e de Montr\'eal.

\appendix
\numberwithin{equation}{section}
\numberwithin{figure}{section}

\section{Comparison of Eq.~\eqref{eq:AHESummary} to the literature}
\textit{It is important to emphasize that the central results of
this paper given in Eq.~\eqref{eq:AHESummary} 
involve the full Green's functions which depend on the self energy $\Sigma(\omega, T)$.} 
Expressions for the anomalous transport coefficients in the 
presence of inelastic scattering allow us to address finite temperatures. 
As far as we are aware, none of these equations have appeared in the literature. 
However, when we drop $\Sigma(\omega, T)$
our results for the Hall conductivity~\cite{Smrcka1977,Bastin1971}, for the
Nernst coefficient~\cite{Gusynin2014,Xiao2006}, and for
the thermal Hall conductivity~\cite{Qin2011,Matsumoto2014,Sharma2016} can be found elsewhere. 
What is also important to emphasize is that, without dynamical self-energy effects, there are many equivalent ways of rewriting these
Green's function-based equations, using different schemes to do a partial integration over frequency.
Notably, however, this equivalence disappears when $\Sigma(\omega, T)$ 
is introduced since the operation of inserting $\Sigma(\omega, T)$ in Green's functions
in general does not commute with an integration by parts over $\omega$.
Thus, only certain representations of these equations are correct, in the presence of inelastic dissipation.

We also emphasize that  Eq.~\eqref{eq:AHESummary} incorporates the proper
subtraction of the divergenceless charge and heat magnetization currents.
In our work we, thereby, derive expressions for the spontaneous orbital charge and heat magnetizations~\cite{Ceresoli2006,Shi2007,Bianco2013,Resta2018}, $\vec{M}^e$ and $\vec{M}^h$,
in the presence of $\Sigma(\omega,T)$, given in the following Eqs.~\eqref{eq:emagnetization} and ~\eqref{eq:hmagnetization}.
We note that, in the absence of $\Sigma(\omega,T)$, an expression for $\vec{M}^e$ 
in terms of Berry curvature (see Eq.~\eqref{eq:MeMh}) is well established in the literature
while that of $\vec{M}^h$ is still controversial~\cite{Qin2011,Xiao2020,Kapustin2020}. 
However, establishing the effects of including $\Sigma(\omega,T)$ on both $\vec{M}^e$ and $\vec{M}^h$ is new to this paper.

\section{Model Hamiltonians}
\label{app:model}

In Sec.~\ref{sec:numerics} of the main text we have looked at two different two-band Hamiltonians on a square lattice,
of the form of $H(\vec{k})$ in Eq.~\eqref{eq:modelHk}.
The model we call $\HI$ derives from a two-band system which supports an anomalous Hall effect; 
it is associated with ferromagnetic metals with active $\{ d_{xz}, d_{yz} \}$ orbitals~\cite{Kontani2007}. 
Written in the form as in Eq.~\eqref{eq:modelHk}, $\HI$ corresponds to
\begin{subequations}
\begin{align}
h_0(\vec{k}) & =-t (\cos k_x + \cos k_y),  \\
h_x(\vec{k}) & = 4 t^\prime \sin k_x \sin k_y , \\
h_y(\vec{k}) & = -\soc,  \\
h_z(\vec{k}) & = - t (\cos k_x -\cos k_y). 
\end{align}
\end{subequations}
Here $\soc$ is the strength of the spin-orbit coupling. $t,t^\prime$ are tight-binding hopping parameters. 

The second Hamiltonian we study, called $\HII$, is derived from a model that has been used to construct flat bands with nonzero Chern number 
and to investigate the possibility of a fractional Chern insulator~\cite{Neupert2011}.
The same model has also been recently exploited to understand the interplay of quantum geometry 
and superconducting fluctuations in the superfluid phase stiffness~\cite{Wang2020,Hofmann2020}. 
For this model we take $\HII(\vect{k})=h_0(\vect{k}) +\vec{h}(\vec{k}) \cdot \boldsymbol{\sigma} -\mu_{\mathrm{F}}$, with
\begin{subequations} \label{eq:HErez}
\begin{align}
  h_0(\vec{k})  &=0,    \label{eq: h0} \\
  h_z(\vec{k})  &  = - 2 t_2 \,  \big[ \cos(k_x + k_y) - \cos(k_x-k_y) \big],    \label{eq: hz} \\
  h_x(\vec{k}) & = -2 t \big[ \cos(\phi+k_y) \cos k_y + \cos(\phi-k_y) \cos k_x \big], \\
  h_y(\vec{k}) & =  2 t \big[ \sin(\phi+k_y) \cos k_y - \sin(\phi-k_y) \cos k_x \big]. 
\end{align}
\end{subequations}
Here,  $t$ and $t_2$ denote the magnitudes of the nearest-neighbor and the second-nearest-neighbor hopping integrals and  
$\phi$ describes a phase associated with the nearest-neighbor hopping. In the original model, the sign of $\phi$ depends on the spin.
For the spin-$\downarrow$ component we consider, $\phi=-\pi/4$.  

The two energy bands of $\HII$ carry opposite Chern numbers, $C_{-}=-1$ ($C_{+}=+1$) for the lower (upper) band.
This can be also seen from Fig.~\ref{fig:WFBerry} which shows that the Berry-curvature distribution of the lower band is dominated by 
negative contributions. For our calculation we use the following definitions for the Berry connection $\boldsymbol{\mathcal{A}}_n(\vec{k}) $, 
Berry curvature $\boldsymbol{\Omega}_n(\vec{k})$, and Chern number $C_n$~\cite{Nagaosa2010}:
\begin{subequations}
\begin{align}
\boldsymbol{\mathcal{A}}_n(\vec{k})   & = -  i \langle u_n | \nabla_{\vec{k}} |u_n\rangle,  \label{eq:ADef}\\
\boldsymbol{\Omega}_n(\vec{k})  & = \nabla_{\vec{k}} \times \boldsymbol{\mathcal{A}}_n(\vec{k}), \label{eq:FnzDef} \\
C_n   & = \frac{1}{2\pi} \int d \vec{k} \;  \Omega_n^{z}(\vec{k}),  \label{eq:ChernDef}
\end{align}
\end{subequations}
where $|u_n\rangle$ is the $n$-th band eigenstate. 
The $\int d\vec{k}$ integral is over the first (reduced) Brillouin zone for $\HI$ ($\HII$). 

In Table~\ref{tab:bandparameter} we list the band parameters used for each model considered in this paper. 

\begin{table}[htbp]
    \centering
    \caption{Tight-binding band parameters used for the models considered. }
    \begin{tabular}{c || c}
        \hline\hline
        ~~~~~~~~Hamiltonians~~~~~~~~        & ~~~~~~~~ Band parameters ~~~~~~~   \\
        \hline
        $\HI$                         & $(t,t^\prime, \soc, \mu_F)=(1, 0.1, 0.25, -1.2)$ \\
        \hline
        $\HII$                        & $(t,t_2,\mu_F)=(1, 1/\sqrt{2}, -2.65)$  \\
        \hline\hline
    \end{tabular}
    \label{tab:bandparameter}
\end{table}

\section{Effects of inelastic scattering on the longitudinal WF law}
\label{app:longitudinalWF}

Although the focus of this paper is on the anomalous transport, for comparison we also give a brief
account of the inelastic scattering effects on the longitudinal transport coefficients.
Using our heat-bath approach, we can easily derive the expressions for longitudinal transport:
\begin{equation} 
\label{eq:longitudinal}
\mathcal{L}_{xx}^{\alpha \beta} = \frac{q^{2-\alpha -\beta}}{2 T^{\beta}} \intdkdw  \Tr \left[v_x \mathcal{A}  v_x \mathcal{A}\right] \omega^{\alpha +\beta} n_F^{(1)}. 
\end{equation}
Here, again, we have put all longitudinal transport coefficients in a consolidated equation.  To be specific, $\{ \mathcal{L}_{xx}^{00}, \mathcal{L}_{xx}^{01}, \mathcal{L}_{xx}^{10}, \mathcal{L}_{xx}^{11}\}
=\{\sigma_{xx}, \beta_{xx},\gamma_{xx},\kappa_{xx}\}$.

\begin{figure}[tp]
\begin{center}
\includegraphics[width=1.\linewidth]{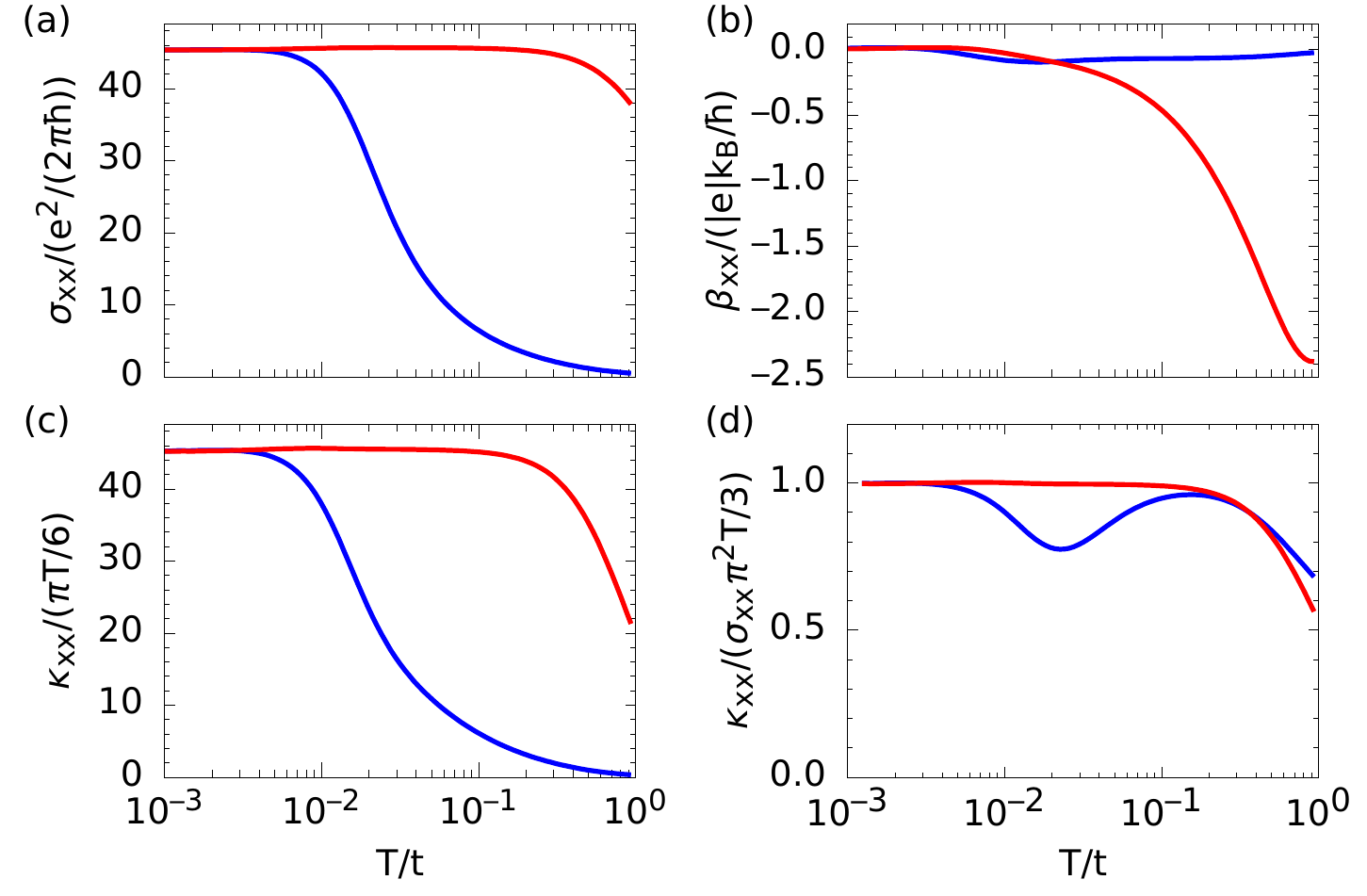} 
\caption{Longitudinal transport coefficients and Wiedemann-Franz ratio, both in the presence (blue) and absence (red) of the inelastic self-energy effects. 
For illustration we use a simple single-band tight-binding model
with the energy dispersion $H(\vec{k}) = - 2 t(\cos k_x + \cos k_y) - \mu_F$ and $\mu_F/t= -1$. 
For the inelastic self energy calculation, we have chosen the electron-phonon coupling strength $\lambda=1$ and the Debye frequency $\omega_D=0.1 t$.
A small residual impurity scattering rate $\sim 0.05 t$ has been included to regularize $\sigma_{xx}$ and $\kappa_{xx}$ so that they do not diverge at $T=0$. }
\label{fig:Longitudinal2}
\end{center}
\end{figure}

Figure~\ref{fig:Longitudinal2} presents the corresponding numerical results for a simple square lattice
both with and without the inelastic self energy $\Sigma$ given in Eq.~\eqref{eq:ImSigmaDef}. 
Quite generally, the inclusion of $\Sigma$ suppresses the magnitude of
the longitudinal transport coefficients at low temperatures $T \lesssim \omega_D \ll T_F$, leading to a
violation of the WF law~\cite{Lavasani2019}.

\section{Numerical results of transport coefficients for the model $\HII$}
\label{app:additionalnumerics}

\begin{figure}[tp]
\includegraphics[width=\linewidth]{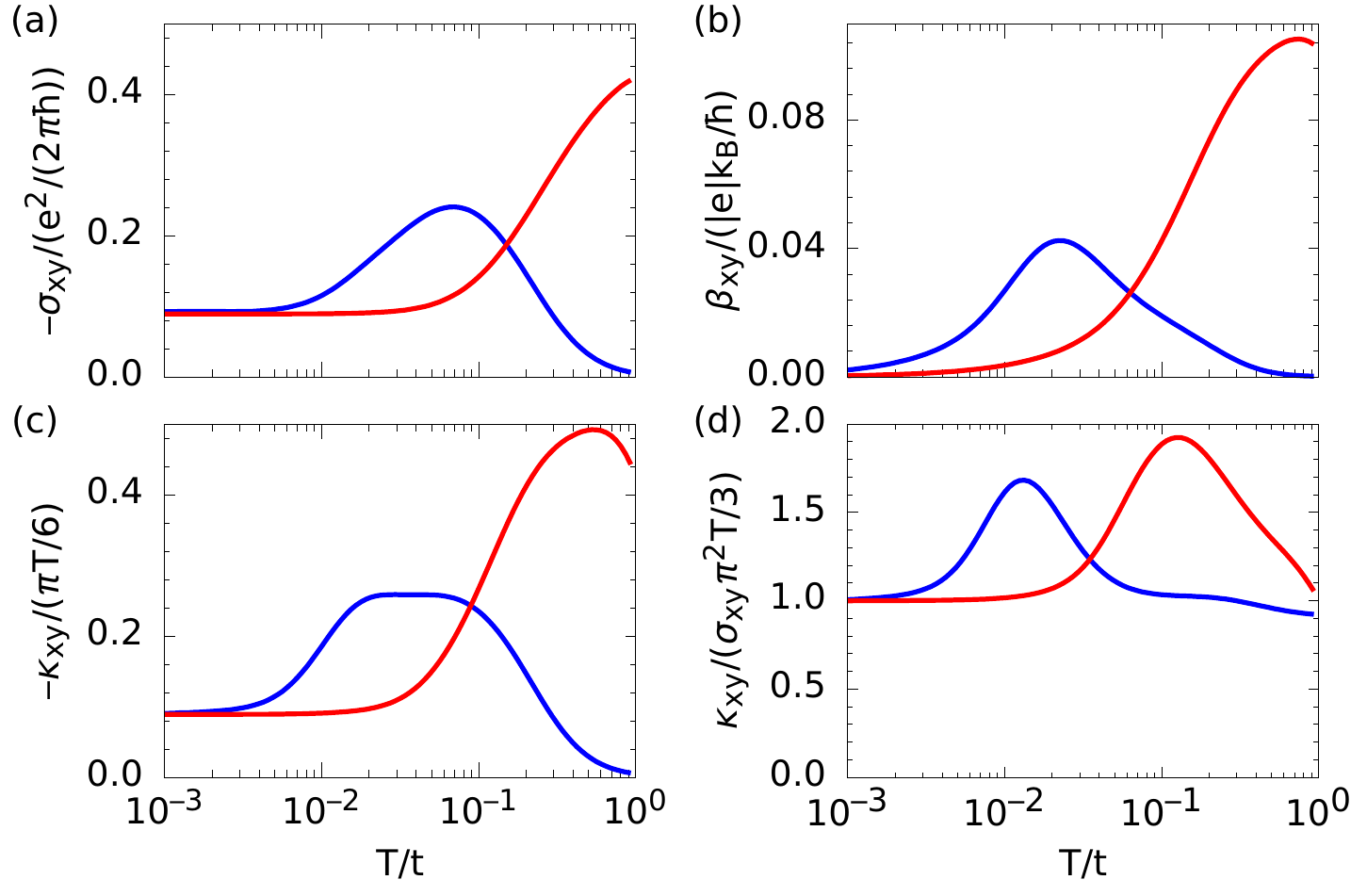} 
\caption{Numerical results for model $\HII$, 
both with (in blue) and without (in red) the inelastic electron self energy effect. 
Here we use the parameters $\omega_D=0.1 t$, $\lambda=5$.  }
\label{fig:ErezTransverseWF1}
\end{figure}

In Fig.~\ref{fig:WFBerry} of the main text we have presented our numerical results for the 
transport ratios
in the case of $\HII$,
although the individual results for each individual transport coefficient were
not shown. They are presented in Fig.~\ref{fig:ErezTransverseWF1}. 
The results for $\sigma_{xy}, \beta_{xy}$ and $\kappa_{xy}$ in the zero-$T$ limit, where the electron-phonon scattering can be neglected,
agree with Eq.~\eqref{eq:cleansigmaxy}. Comparing Fig.~\ref{fig:ErezTransverseWF1} to Fig.~\ref{fig:Longitudinal2} we see that the temperature dependence of
the anomalous transport coefficients can be much more complex than that of their longitudinal counterparts. 

\section{Derivation of spontaneous orbital charge and heat magnetizations}
\label{app:magnetization}

In this appendix, we derive the charge and heat (orbital) magnetizations
associated with Berry-curvature effects: $\vec{M}^e$ and $\vec{M}^h$. These are
obtained by relating them to charge and heat boundary currents.
To calculate the latter, we imagine the system in the $xy$-plane with a boundary. 
At local equilibrium we assume the system's local chemical potential varies slowly from its bulk value $\mu_F$ to its vacuum value $\mu_F=- \infty$ at the boundary edge.
In addition, the electric potential created by the boundary is also 
experienced by the bath particles since they carry the same charge as those in the system. 
Using the local-equilibrium assumption, in response to this potential, the bath particle energy level $a_{\alpha}$ in Eq.~\eqref{eq:Hb} adjusts itself together with
$\mu_F$ so that their difference remains a constant across the whole system; in this way
the source field $D_i$ in Eq.~\eqref{eq:psimotion} has an indirect dependence on $\mu_F$. 

Within a linear expansion, the above boundary effects can be taken into account by making the following substitution in the EOM in Eq.~\eqref{eq:psimotion}:
\begin{subequations} \label{eq:nablamuF}
\begin{align}
\mu_F     & \rightarrow \mu_F + \nabla \mu_F \cdot \vec{x}_i, \\
\Sigma    &  \rightarrow \Sigma +  \frac{\partial \Sigma}{\partial \mu_F} \nabla \mu_F \cdot \vec{x}_i, \\
D_i  & \rightarrow D_i + \frac{\partial D_i}{\partial \mu_F} \nabla \mu_F \cdot \vec{x}_i, 
\end{align}
\end{subequations}
where $\nabla \mu_F$ is a constant. We emphasize that it is important to consider simultaneously the $\mu_F$ dependence of both $\Sigma$ and $D_i$
in order to ensure that the fluctuation-dissipation relation, Eq.~\eqref{eq:FDT}, holds across the whole system,
reflecting our assumption of local equilibrium. 

The calculation of the boundary charge current density, $\langle \vec{J}^e \rangle_{\nabla \mu_F} $, 
due to the perturbations in Eq.~\eqref{eq:nablamuF}, closely follows that
of $\langle \vec{J}^h \rangle_{\gradT}$ in Sec.~\ref{sec:kappaxyderivation} of the main text:
\begin{widetext}
\begin{align}
\langle \vec{J}^e \rangle_{\nabla \mu_F} &= \frac{q}{i} \intdkdw  \Tr \left[ \vec{v}(\vec{k}) \Glesser_{\nabla \mu_F}(\vec{k},\omega) \right]        \nonumber  \\
  & = - i q  \intdkdw  \Tr  \bigg \{ \vec{v} G_\R  \left(1- \frac{\partial \Sigma_\R}{\partial \mu_F}\right) \partial_j \left[ G_\R \langle D^\dagger D\rangle G_\A \right]  \nonumber \\
  & \quad  + \vec{v} G_\R \langle D^\dagger D \rangle G_\A \left(1- \frac{\partial \Sigma_\A}{\partial \mu_F}\right)  \partial_j  G_{\A}  - \vec{v}  G_\R    \left[  \frac{\partial }{\partial \mu_F} \langle D^\dagger D \rangle \right] \partial_j G_\A \bigg\}  (\nabla \mu_F)_j   \nonumber \\
 &   = q  \intdkodw  \mathrm{Tr} \left[   \vec{v} G  \left(1- \frac{\partial \Sigma}{\partial \mu_F}\right) \partial_j  G  \right] n_F  \;   (\nabla \mu_F)_j . 
\end{align}
\end{widetext}
The subscript $\nabla \mu_F$ in $\Glesser_{\nabla \mu_F}$ indicates the latter is the perturbed lesser Green's function. 
When the subscript is not specified this refers to the bulk.
To arrive at the last line in the equation above,
we have used the fluctuation-dissipation relation, Eq.~\eqref{eq:FDT}. 
Here, as before, the integral $\oint$ is along the contour in Fig.~\ref{fig:contour}. 

From $\langle \vec{J}^e \rangle_{\nabla \mu_F}$ we calculate the charge magnetization by using~\cite{Cooper1997}:
\begin{gather}
\langle \vec{J}^e \rangle_{\nabla \mu_F} = \nabla \mu_F \times \frac{\partial \vec{M}^e}{\partial \mu_F} , 
\end{gather}
which leads to 
\begin{align} \label{eq:dMdT}
 \frac{\partial M_z^e}{\partial \mu_F} 
& =  - q  \frac{\epsilon_{zij}}{2} \ointdw  \Tr [ v_i  ( \partial_\omega G ) v_j G] n_F(\omega)  \nonumber\\ 
& =  -q  \frac{\epsilon_{zij}}{2}    \oint \frac{d \bar{\omega}}{2\pi}  \Tr[ v_i  ( \partial_{\bar{\omega}} G ) v_j G] n_F(\bar{\omega}-\mu_F) \nonumber\\ 
& =  \frac{\partial}{\partial \mu_F} \bigg\{ q  \frac{\epsilon_{zij}}{2} \oint \frac{d \bar{\omega}}{2\pi}  \mathrm{Tr}[ v_i ( \partial_{\bar{\omega}} G) v_j G]  \nonumber\\
&\quad\times\frac{1}{\beta} \mathrm{Li}_1\left(-e^{-\beta(\bar{\omega} -\mu_F)}\right)  \bigg\}. 
\end{align}
For brevity we have suppressed $\intdk$ in these equations. 
In arriving at the first line we have used $G(1-\partial_{\mu_F}\Sigma) G = G(1-\partial_{\omega}\Sigma) G = -\partial_\omega G$ since $\Sigma$ needs to be a function of $\omega+\mu_F$ 
(i.~e., it cannot depend on the choice of energy reference point). 
From the first to the second line we have changed variables $\omega \rightarrow \bar{\omega}=\omega + \mu_F$. 
The last line is obtained by using $n_F(\omega)= (1/\beta) \partial_{\omega} \mathrm{Li}_1(-e^{-\beta \omega})$. 

Integrating $\partial M_z^e/\partial \mu_F$ over $\mu_{F}$ (from its vacuum to its bulk value) gives 
\begin{align} \label{eq:emagnetization}
 M_z^e & = q  \frac{\epsilon_{zij}}{2} \intdkodw  \mathrm{Tr}\left[ v_i \partial_\omega G v_j G\right]  \frac{1}{\beta} \mathrm{Li}_1(-e^{-\beta \omega}),
\end{align} 
where we have changed the integration variable $\bar{\omega}$ back to $\omega$ and restored $\intdk$. 

The derivation of the heat magnetization follows closely that of $M_z^e$. The expression for $\partial M_z^h/\partial \mu_F$ differs from
that of $\partial M_z^e/\partial \mu_F$ in the first line of Eq.~\eqref{eq:dMdT} only by a factor of $\omega/q$, i.~e.,
\begin{widetext}
\begin{align}
\frac{\partial M_z^h}{\partial \mu_F}  & = -   \frac{\epsilon_{zij}}{2} \intdkodw \Tr  [ \omega v_i  ( \partial_\omega G)  v_j G] n_F(\omega) \nonumber\\
& = \frac{\partial}{\partial \mu_F}  \bigg\{ -  \frac{\epsilon_{zij}}{2} \intdk \oint \frac{d \bar{\omega}}{2\pi}  \mathrm{Tr} [ v_i (\partial_{\bar{\omega}} G)  v_j G] 
Q( \bar{\omega} -\mu_F) \bigg\}, 
\end{align}
\end{widetext}
where $Q$ is defined in Eq.~\eqref{eq:Qdef} of the main text and reproduced here for convenience, 
\begin{equation}
Q(\omega) \equiv  \frac{\omega}{\beta}  \mathrm{Li}_1 (- e^{-\beta \omega}) 
+ \frac{1}{\beta^2}\mathrm{Li}_2(- e^{-\beta \omega}). 
\end{equation}
Integrating $\partial M_z^h / \partial \mu_F$ over $\mu_F$ and changing $\bar{\omega}$ back to $\omega$ leads to 
\begin{align}
M_z^h   =   \frac{\epsilon_{zij}}{2} \intdkodw  \Tr [  v_i   ( \partial_{\omega} G)  v_j G]  Q(\omega).     \label{eq:hmagnetization}
\end{align}

In the absence of $\Sigma(\omega,T)$, one can compute the $\omega$ integrals in Eqs.~\eqref{eq:emagnetization}
and \eqref{eq:hmagnetization} and obtain 
\begin{subequations} \label{eq:MeMh}
\begin{align}
\vec{M}^e & = \sum_n \int \frac{d\vec{k}}{(2\pi)^2}  \bigg[ q \vec{m}_n   n_F(\xi_n) 
- q \boldsymbol{\Omega}_n \frac{1}{\beta} \mathrm{Li}_1(- e^{-\beta \xi_n}) \bigg], \label{eq:MeWOSigma} \\
\vec{M}^h & =  \sum_n \int \frac{d\vec{k}}{(2\pi)^2}  \bigg[ \xi_n \vec{m}_n  n_F(\xi_n) - \boldsymbol{\Omega}_n Q(\xi_n) \bigg], 
\label{eq:MhWOSigma}
\end{align}
\end{subequations}
where 
\begin{equation}
\vec{m}_n   \equiv  \frac{i }{2} \langle \nabla_{\vec{k}} u_{n}\left| \times [H -  \xi_n] \right| \nabla_{\vec{k}} u_n \rangle. 
\end{equation}
Both $\vec{M}^e$ and $\vec{M}^h$ contain two parts. The first part involves $q \vec{m}_n$ and $\xi_n \vec{m}_n$, which can be interpreted as charge and heat orbital magnetic moments of the $n$-th eigenstate, respectively~\cite{Xiao2006,Shi2007,Zhang2016,Xiao2020}. 
In a semiclassical picture, these orbital magnetic moments originate from self-rotation of a wavepacket~\cite{Xiao2006,Zhang2016}. 
The other part of $\vec{M}^e$ and $\vec{M}^h$ involves the band-dependent Berry curvature,  $\boldsymbol{\Omega}_n (\vec{k}) $, whose definition is given in Eq.~\eqref{eq:FnzDef}. 
This part is associated with the center of mass motion of a wavepacket in a semiclassical theory.
We note that, in Eq.~\eqref{eq:MeMh}, we have written the magnetizations in their fully vectorial form to indicate that they
also apply to three dimensions as well. 

When compared to the literature we find our expression for $\vec{M}^e$ in Eq.~\eqref{eq:MeWOSigma} agrees with results in Refs.~\cite{Xiao2006,Shi2007,Bianco2013}. 
We also find that our expression for the heat magnetization
$\vec{M}^h$ in Eq.~\eqref{eq:MhWOSigma} coincides with results in Refs.~\cite{Shitade2014,Zhang2016,Xiao2020}. 
However, when comparing with $\vec{M}^h$ derived in Ref.~\cite{Qin2011}, 
we find that Ref.~\cite{Qin2011} contains an extra term that involves $n_F$,
which arises because of a particular scaling form of the heat-current operator, Eq.~(5) of Ref.~\cite{Qin2011}, has been assumed. 
However, this scaling relation does not hold on a lattice~\cite{Kapustin2021}. Therefore, the $\vec{M}^h$ result in Ref.~\cite{Qin2011} cannot be directly applied to a lattice model such as ours.

\section{Derivations of anomalous $\sigma_{xy}, \gamma_{xy}$ and $\beta_{xy}$}
\label{app:sigmagammabeta}

In the main text we have derived $\kappa_{xy}$ using our heat-bath approach. 
In this section we provide detailed derivations of the other three anomalous transport coefficients, $\{\sigma_{xy}, \gamma_{xy},\beta_{xy}\}$. 

\subsection{Derivation of $\sigma_{xy}$}
We start with the response to the electric field $\vec{E}$. The perturbation due to $\vec{E}$ can be written (in $\vec{k}$ space)as
\begin{gather}
H_{\vec{E}}^\prime(\vec{k}) =  q \vec{v}(\vec{k}) \cdot \vec{E} t, 
\end{gather}
where we have adopted the temporal gauge $\vec{A}_{\vec{E}} = - \vec{E} t$, where $t$ is the time~\cite{Tan2004}. 
It should be understood that $H_{\vec{E}}^\prime$ is a matrix in orbital subspace. 
Note that $\vec{A}_{\vec{E}}$ does not affect the heat-bath degrees of freedom because they are local and do not couple to
the potential $\vec{A}_{\vec{E}}$. 

The electric current density response, $\langle \vec{J}^e \rangle_{\vec{E}}$, due to $H_{\vec{E}}^\prime$ can be computed similarly as
$\langle \vec{J}^e \rangle_{\nabla \mu_F}$ in Appendix~\ref{app:magnetization}. 
\begin{widetext}
\begin{align} \label{eq:JvecE}
\langle \vec{J}^e \rangle_{\vec{E}}  
&  = q  \frac{1}{i} \intdkdw \Tr [ \vec{v} \Glesser_{\vec{E}}]   \nonumber \\
& = q  \intdkdw \Tr \bigg\{ \vec{v} \bigg[   G_\R H_{\vec{E}}^\prime  G_\R \langle D^\dagger D \rangle G_\A  +  G_\R  \langle D^\dagger D \rangle G_\A  H_{\vec{E}}^\prime  G_\A  \bigg]  \bigg\}   \nonumber  \\
&=  - q^2  \intdkdw    \Tr  \bigg\{  \vec{v} \bigg[ G_\R  v_j  (\partial_\omega \Glesser  )    +  \Glesser v_j (\partial_\omega G_\A )  \bigg]   \bigg\} E_j  \nonumber \\
& =  i  q^2  \intdkdw    \Tr \bigg\{ \vec{v}  \bigg[  (\partial_\omega G_\R) v_j \mathcal{A}   -   \mathcal{A}  v_j  ( \partial_\omega  G_\A) \bigg]  \bigg\} n_F  E_j . 
\end{align}
\end{widetext}
We have again used $t=-i \partial_\omega$, the fluctuation-dissipation relation Eq.~\eqref{eq:FDT}, and the definition $\Glesser=  (G_\A -G_\R) n_F \equiv i  \mathcal{A} n_F$ in Eq.~\eqref{eq:Glesser0}. 
From Eq.~\eqref{eq:JvecE} we obtain 
\begin{widetext}
\begin{align} \label{eq:Bastin}
\sigma_{ij} =  i  q^2  \frac{\epsilon_{zij} \epsilon_{zab}}{2}  \intdkdw  \Tr \bigg[  v_a   (\partial_\omega G_\R) v_b  \mathcal{A} 
-  v_a  \mathcal{A} v_b ( \partial_\omega G_\A)  \bigg ]  n_F. 
\end{align}
\end{widetext}
The antisymmetrization factor $\epsilon_{zij} \epsilon_{zab} / 2$ is included since only the transverse part contributes to anomalous Hall transport. 
Equation~\eqref{eq:Bastin} is in the well known Bastin form~\cite{Bastin1971,Crepieux2001}. 
However, we emphasize that our Eq.~\eqref{eq:Bastin} has the inelastic self energy naturally built in. 
One can also rewrite $\sigma_{ij}$ in a Streda-like form~\cite{Crepieux2001} by splitting $\sigma_{ij}$ into two
halves and performing an integration by parts over $\omega$ on the second half,
leading to $\mathcal{L}_{ij}^{00}$ in Eq.~\eqref{eq:AHESummary}.

\subsection{Derivation of $\gamma_{xy}$}

Next we derive the anomalous $\gamma_{xy}$ coefficient, which describes a heat-current response to the applied electric field. 
Unlike in the previous case, a calculation of $\gamma_{xy}$ requires a proper subtraction of the charge magnetization current contribution (see Eq.~\eqref{eq:Msubtraction}). 
To derive $\gamma_{xy}$ we first calculate the ``microscopic'' heat current density due to the perturbation $H_{\vec{E}}^\prime$, 
which follows closely that of $\langle \vec{J}^h \rangle_{\gradT}$ in Sec.~\ref{sec:kappaxyderivation} of the main text. 
\begin{widetext}
\begin{align}
\langle \vec{J}^h \rangle_{\vec{E}} & =\frac{1}{i} \frac{1}{V} \int d\vec{x} dt  \int d \vec{x}^\prime d t^\prime \Tr \left[   \vec{v}(\vec{x}^\prime, \vec{x})  \frac{ i   \partial_t   - i  \partial_{t^\prime} }{2} \Glesser_{\vec{E}}(\vec{x} t, \vec{x}^\prime t^\prime)  \right]   \nonumber \\
& =  \frac{1}{2}  \intdkdw  \Tr \bigg \{  \vec{v} \omega \bigg[  G_\R  H_{\vec{E}}^\prime G_\R \langle D^\dagger D \rangle G_\A  +  G_\R  \langle D^\dagger D \rangle G_\A  H_{\vec{E}}^\prime  G_\A \bigg]  \nonumber \\
& \quad +   \vec{v}  \left[  G_\R  H_{\vec{E}}^\prime G_\R \langle D^\dagger D \rangle \omega G_\A  +  G_\R  \langle D^\dagger D \rangle G_\A  H_{\vec{E}}^\prime  \omega G_\A  \right] \bigg \}   \nonumber \\
& =  - i  \frac{q}{2}    \intdkdw  \Tr \left\{  \vec{v}  \omega  \bigg[  G_\R v_j  \partial_\omega ( \mathcal{A}  n_F) +  \mathcal{A} n_F v_j \partial_\omega G_\A \bigg]  
 - \vec{v}  \left[  (\partial_\omega G_\R)  v_j  \omega  \mathcal{A}  n_F +  [\partial_\omega ( \mathcal{A} n_F)] v_j   \omega G_\A  \right]  \right\}  E_j . 
\end{align}
%\end{widetext}
This leads to $\langle J_i^h \rangle_{\vec{E}} = L_{ij}^{h,e} E_j$, with  
%\begin{widetext}
\begin{align}
L_{ij}^{h,e} & = \frac{  \epsilon_{zij} \epsilon_{zab} }{2} q \intdk   \bigg\{ \intdw \Tr \big[ v_a  G_\R v_b G_\A \big]   \omega (- \partial_\omega n_F)
 -  \ointdw  \mathrm{Tr} \big[  v_a ( \partial_\omega G) v_b G \big ]   \omega  n_F \bigg\}. 
\end{align}
\end{widetext}
 
According to Eq.~\eqref{eq:Currents}, the transport thermoelectric coefficient $\gamma_{ij}$ is given by
\begin{align}
\gamma_{ij}  &  = L_{ij}^{h,e} - \epsilon_{zji} M^e_z. 
\end{align}
Using Eq.~\eqref{eq:emagnetization} for $M^e_z$, we immediately see 
that $\gamma_{ij}=\mathcal{L}_{ij}^{10}$ in Eq.~\eqref{eq:AHESummary} of the main text.

\subsection{Derivation of $\beta_{xy}$}

Now we present the derivation of the transport charge current response to an applied temperature gradient $\gradT=-\nabla T$. 
This derivation follows that of $\kappa_{xy}$ in Sec.~\ref{sec:kappaxyderivation} in
large part. Without repeating the details
we note $\langle J^e_i \rangle_{\gradT} \equiv L_{ij}^{e,h} (-\partial_j T)$ with $L_{ij}^{e,h}$
differing from that of $L_{ij}^{h,h}$ in Eq.~\eqref{eq:Lhh} by a factor of $q/\omega$, i.~e., 
%\begin{widetext}
\begin{multline} \label{eq:Leh}
L_{ij}^{e,h}   =  -  \frac{\epsilon_{zij} \epsilon_{zab}}{2} q \intdkdw \Tr \bigg\{ v_a 
\bigg[   G_\R    \frac{\partial \Sigma_\R }{\partial T} \partial_{b} \Glesser  \\
+ \Glesser \frac{\partial \Sigma_\A}{\partial T} \partial_{b}  G_\A 
+  i G_\R    \left(  \frac{\partial}{\partial T} \langle  D^\dagger D \rangle \right)   \partial_{b} G_\A
 \bigg] \bigg\}. 
\end{multline}
%\end{widetext}
To obtain the transport $\beta_{xy}$ we again need to subtract the magnetization current contribution.
From Eq.~\eqref{eq:Currents} 
\begin{align}
\beta_{ij}  & =  L_{ij}^{e,h} - \epsilon_{izj} \frac{\partial M_z^e}{\partial T}  \equiv \beta_{ij}^{(1)} + \beta_{ij}^{(2)}. 
\label{eq:betaxysplit}
\end{align}
As in Eq.~\eqref{eq:kappaxysplit} we divide $\beta_{ij}$ into two pieces. $\beta_{ij}^{(1)}$ contains all contributions
that involve $\partial \Sigma/\partial T$, while $\beta_{ij}^{(2)}$ contains the remainder. 
Now we show that, as a consequence of the fluctuation-dissipation relation, Eq.~\eqref{eq:FDT}, $\beta_{ij}^{(1)} \equiv 0$. 
From Eq.~\eqref{eq:emagnetization} we obtain
\begin{widetext}
\begin{align} \label{eq:dMedT}
\frac{\partial M^e_z}{\partial T}   
& = q  \frac{\epsilon_{zij}}{2} \intdkodw   \Tr \bigg\{  \partial_\omega \bigg[   v_i  (\partial_T G)  v_j G  \bigg]  \frac{1}{\beta} \mathrm{Li}_1(-e^{-\beta \omega})
 +  v_i  ( \partial_\omega G)  v_j G   \bigg[\mathrm{Li}_1(-e^{-\beta \omega}) - \beta \omega n_F \bigg]  \bigg\}.   
\end{align}
\end{widetext}
In this equation, the first term depends on $\partial \Sigma/\partial T$ while the second does not. 
Using Eqs.~\eqref{eq:Leh} and \eqref{eq:dMedT} in Eq.~\eqref{eq:betaxysplit} and collecting all $\partial \Sigma/\partial T$ terms leads to 
\begin{widetext}
\begin{align}
\beta_{ij}^{(1)}   &  =   -   \frac{\epsilon_{zij} \epsilon_{zab}}{2}  \intdkdw
 \Tr \bigg [  v_a  G_\R \frac{\partial \Sigma_{\R} }{\partial T} \partial_b \Glesser 
 + v_a  \Glesser \frac{ \partial \Sigma_{\A} }{\partial T}  \partial_b G_{\A}
 +  v_a   G_{\R} \frac{ \partial (- 2 i \ImSigma)}{\partial T} \partial_b G_{\A}   n_F \bigg] \nonumber \\
 & \quad +   \frac{\epsilon_{zij} \epsilon_{zab}}{2}  \intdkodw  \bigg( \partial_\omega \Tr \big[   v_a  (\partial_T G)  v_b G   \big] \bigg) \frac{1}{\beta} \mathrm{Li}_1(-e^{-\beta \omega}) \nonumber\\ 
 &  =   \frac{\epsilon_{zij} \epsilon_{zab}}{2}  \intdkodw  \bigg\{   \Tr \big[ v_a  (\partial_T G)  v_b G  \big]   n_F
  -  \Tr \big[  v_a   (\partial_T   G)  v_b G  \big]  \partial_\omega   \frac{1}{\beta} \mathrm{Li}_1(-e^{-\beta \omega})
   \bigg\}  \nonumber \\
 & =0. 
\end{align}
\end{widetext}
To obtain the last line we have used $\mathrm{Li}_1(x)= - \ln(1-x)$. 
Again, we emphasize that the fluctuation-dissipation relation in Eq.~\eqref{eq:FDT} is essential for the complete cancellation above. 
What remains in Eq.~\eqref{eq:betaxysplit} is then 
\begin{widetext}
\begin{align}
\beta_{ij}^{(2)}  
& =
 q  \frac{\epsilon_{zij} \epsilon_{zab}}{2} \intdk 
 \bigg\{ \intdw  \Tr[v_a G_\R v_b G_\A ] \left(-\frac{\omega}{T} \partial_\omega n_F \right) 
  + \ointdw \Tr [ v_a  (\partial_\omega G) v_b G ]  \big[\mathrm{Li}_1(-e^{-\beta \omega}) - \beta \omega n_F \big]  \bigg\}, 
\end{align}
\end{widetext}
which leads directly to $\mathcal{L}_{ij}^{01}$ in Eq.~\eqref{eq:AHESummary} of the main text.

\section{Current operators and continuity equations}
\label{app:currentdef}

In this section we define the charge and heat-current operators that we use to compute the transport coefficients 
and show they satisfy  two continuity equations. 
We first define the charge and energy densities ($\rho_i^e$ and $\rho_i^h$) at lattice site $i$ from Eq.~\eqref{eq:HsHbHsb} as
\begin{widetext}
\begin{subequations}
\begin{align}\label{eq:rhoe}
\rho_i^e  & \equiv q\big[  \sum_{m} \psi_{im}^\dagger\psi_{im}+  \sum_\alpha  \phi_{i\alpha}^\dagger \phi_{i\alpha} \big],  \\
\rho_i^h  & \equiv  \frac{1}{2}  \sum_{j} \sum_{mn} [ \psi_{im}^\dagger H_{im, jn} \psi_{jn} + h.c.]  -\mu_F \sum_m \psi^\dagger_{im} \psi_{im} + \sum_\alpha a_\alpha \phi^\dagger_{i\alpha} \psi_{i\alpha} 
+ \sum_{\alpha m} \eta_{\alpha} [\psi^\dagger_{im} \phi_{i\alpha} + \mathrm{h.c.}  ] .
\end{align}
\end{subequations}
\end{widetext}
In the first term of $\rho_i^h$, the factor of $1/2$ is needed to avoid double counting so that the total energy, i.~e., the Hamiltonian, satisfies $H= \sum_i \rho_i^h$. 
For brevity we have suppressed the time dependence of operators on both sides of these equations. 
Next we define the corresponding charge and heat-current density operators on the link that connects lattice site $i$ and $j$ as~\cite{Paul2003}
\begin{widetext}
\begin{subequations} \label{eq:JeJh}
\begin{align}
\vect{J}_{ij}^e   & \equiv   \frac{q}{2i} \sum_{mn}  \big[  \psi_{im}^\dagger H_{im,jn}  (\vec{x}_i -\vec{x}_j)    \psi_{jn}  - \mathrm{ h.c. } \big] , \\
\vect{J}_{ij}^h & \equiv  \frac{1}{4} \sum_{mn}  
\bigg\{ 
\bigg[
 \psi_{im}^\dagger H_{im,jn}  (\vec{x}_i -\vec{x}_j)   \dot{\psi}_{jn}  
- \dot{\psi}_{im}^\dagger H_{im,jn}  (\vec{x}_i -\vec{x}_j)  \psi_{jn}   \bigg]
+ \mathrm{h.c.} 
\bigg\},
\label{eq:Jhij}
\end{align}
\end{subequations}
\end{widetext}
where $\dot{\psi}_{i m} \equiv   \partial_t \psi_{im}$ and $\vec{x}_i$ is the position of lattice site $i$. 
The heat-bath fermion field, $\phi_{i\alpha}$, does not enter the expressions of the
charge and heat-current densities because it is local.  

Using the following coupled equations of motion, derived from Eq.~\eqref{eq:HsHbHsb}, 
\begin{subequations}
\begin{align}
i \dot{\psi}_{im}   &   =  \sum_{jn} H_{im,jn} \psi_{jn}  - \mu_F \psi_{im}  + \sum_{\alpha} \eta_{\alpha} \phi_{i\alpha}, \\
i \dot{\phi}_{i\alpha}  & = a_\alpha \phi_{i\alpha} + \sum_{m} \eta_{\alpha}  \psi_{im},
\end{align}
\end{subequations}
one can easily prove that the following two continuity equations are satisfied: 
\begin{subequations} \label{eq:eh_continuity}
\begin{align}
\partial_t \rho_i^e+ \sum_{j\ne i}  \frac{(\vec{x}_j - \vec{x}_i)}{|\vec{x}_j -\vec{x}_i |^2} \cdot (2  \vect{J}_{ij}^e )&=0,\\
\partial_t \rho_i^h+ \sum_{j\ne i}  \frac{(\vec{x}_j - \vec{x}_i)}{|\vec{x}_j-\vec{x}_i |^2} \cdot (2 \vect{J}_{ij}^h ) &=0. 
\label{eq:h_continuity}
\end{align}
\end{subequations}
These are the discretized form of the more familiar continuum equations, $\partial_t \rho^e + \nabla \cdot \vec{J}^e=0$
and $\partial_t \rho^h + \nabla \cdot \vec{J}^h=0$~\cite{Boykin2010}.
The factor $2$ in $2  \vect{J}_{ij}^e $ and $2  \vect{J}_{ij}^h $ of Eq.~\eqref{eq:eh_continuity} appears because
the total current along the link $ij$ is given by $\vect{J}_{ij}^{e/h}+\vect{J}_{ji}^{e/h}= 2 \vect{J}_{ij}^{e/h}$~\cite{Boykin2010}. 

In the main text, for convenience we have used a continuous notation for $\vec{J}^h$ in Eq.~\eqref{eq:JhgradT}, which can be obtained from the discrete version in Eq.~\eqref{eq:Jhij}
by replacing $\{\psi^\dagger_{im}, -i H_{im,jn} (\vec{x}_i -\vec{x}_j), \psi_{jn} \} $ with  $\{\psi^\dagger(\vec{x}^\prime), \vec{v}(\vec{x}^\prime,\vec{x}), \psi(\vec{x}^\prime) \}$. 
Note that $\{\psi^\dagger(\vec{x}^\prime), \vec{v}(\vec{x}^\prime,\vec{x}), \psi(\vec{x}^\prime) \}$ are either vectors or matrices in the orbital subspace. 

% \clearpage
% \bibliography{References}

%apsrev4-2.bst 2019-01-14 (MD) hand-edited version of apsrev4-1.bst
%Control: key (0)
%Control: author (8) initials jnrlst
%Control: editor formatted (1) identically to author
%Control: production of article title (0) allowed
%Control: page (0) single
%Control: year (1) truncated
%Control: production of eprint (0) enabled
%

\end{document}